\documentclass[ams, prb, twocolumn, amssymb,floatfix]{revtex4-1}
\usepackage{amsmath}
\usepackage{tabularx}
\usepackage{bm}
\usepackage{slashed}
\usepackage{euscript}
\usepackage{graphicx}
\usepackage{color}
\usepackage{amsfonts}
\usepackage{exscale}
\usepackage{amsbsy}
\usepackage{subfigure}
\usepackage{textcomp}
\usepackage{comment}
\usepackage{mathrsfs}
\usepackage{bbold}
\usepackage{blkarray}
\usepackage{ulem}

\usepackage{hyperref}

\newcommand{\nn}{\nonumber}

\newcommand{\beq}{\begin{equation}}
\newcommand{\eeq}{\end{equation}}
\newcommand{\be}{\begin{eqnarray}}
\newcommand{\ee}{\end{eqnarray}}

\begin{document}
	
	\title{Interaction effects on quantum Hall transitions: dynamical scaling laws and superuniversality}
	\author{Prashant Kumar$^{1,2}$, P. A. Nosov$^1$, S. Raghu$^{1,3}$}
	\affiliation{$^{1}$Stanford Institute for Theoretical Physics, Stanford University, Stanford, California 94305, USA\\
		$^2$Department of Physics, Princeton University, Princeton NJ 08544, USA\\
		$^{3}$Stanford Institute for Materials and Energy Sciences,\\
		SLAC National Accelerator Laboratory, Menlo Park, CA 94025, USA}
	\date{\today}
	
	\begin{abstract}
		
		We study the role of electron-electron interactions  near integer and abelian fractional quantum Hall (QH) transitions using composite fermion (CF) representations.   Interaction effects are encapsulated in CF theories as gauge fluctuations.  Without gauge fluctuations, the CF system realizes a `dual' representation of the non-interacting QH transition.  With gauge fluctuations, the system is governed by a gauged nonlinear sigma model (NLSM) with a $\theta-$term.
		While the transition is described by a strong-coupling fixed point of the NLSM, we are nevertheless able to deduce two of its properties.   With $1/r$ interactions, 1) the transition has a dynamical exponent $z=1$, and 2) all transitions are `superuniversal': fractional and integer QH transitions are in the same universality class.  With short-range interactions, $z=2$ and the fate of superuniversality remains unclear. 
	\end{abstract}

	\maketitle

	\section{Introduction }
	The quantum Hall to insulator transition (QHIT) is one of the best studied, and archetypal families of metallic quantum critical points\cite{SondhiGirvinCariniShahar}.  Nevertheless, many fundamental issues remain poorly understood.  For instance, a commonly held view is that   the integer QHIT always has a non-interacting description -  afterall,  the phases on either side of the transition have free-fermion prototypes.  By contrast, it seems absurd to suppose that the fractional QHIT may map on to free particles.  Thus, at a first glance the integer and fractional QHITs appear to be completely distinct.  
	
	The flaw with this reasoning is the premise that interactions can be neglected near the integer QHIT: as a point of principle, adiabatic continuity to free fermion ground states breaks down at a critical point.  Moreover, the very fact that the resistivity tensor is finite and universal as $T \rightarrow 0$ cannot be obtained from free electrons in a disordered Landau level, since extended states occur only at one energy\cite{Wang2000}.  Furthermore, there is evidence for dynamical scaling laws, which can only be explained by invoking interactions.\cite{Engel1993,Wei1994,Li2009}  A description of the QHIT including both the effects of disorder and interactions has remained elusive.  
	
	Motivated by these considerations, we have formulated the QHIT problem in a dual composite fermion (CF) representation, where interaction effects are captured by fluctuations of a dynamical $U(1)$ gauge field.  There are two distinct CF theories that can be used to study the QHIT.  The first is the theory of Halperin, Lee and Read (HLR)\cite{Halperin1993}, obtained by attaching two flux quanta to electrons via a singular gauge transformation.  The second, more recent theory due to Son involves Dirac CFs, which can be thought of as dual fermionic vortex degrees of freedom\cite{Son2015}.  When gauge fluctuations are ignored within a mean-field approximation, both CF theories lead to the same predictions at criticality\cite{Kumar2019b,Huang2021}.  	
	\begin{figure}
		\includegraphics[width=2.3in]{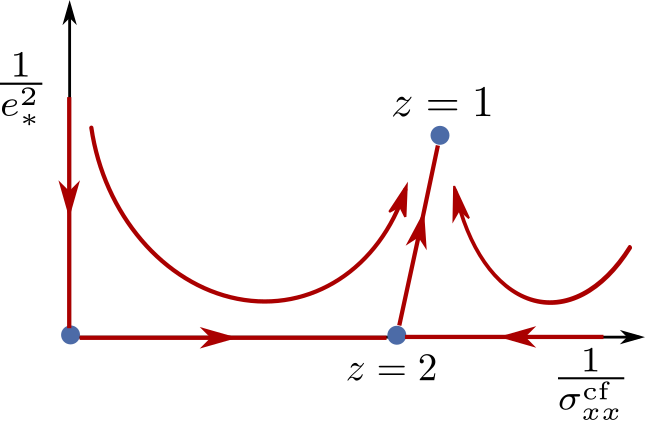}
		\caption{Schematic renormalization group flows at criticality with $U= e_*^2/r$  from the viewpoint of the Dirac composite fermions with a random vector potential.  The Hall conductivity is not renormalized by gauge fluctuations at criticality.  Hence, at criticality there are 2 running couplings, $1/\sigma^{\rm cf}_{xx}$, the disorder strength and $1/e_*^2$, the gauge fluctuation strength. At $1/e_*^2 = 0$, the mean-field CF theory is valid and the theory flows towards the IQHIT fixed point of non-interacting CFs characterized by $z=2$. However, when $1/e_*^2 >0$, it becomes unstable to an interacting fixed point with $z=1$ scaling.
		}
		\label{pd}
	\end{figure} 
	
	Here, we proceed further, and take into account gauge fluctuations in the presence of quenched randomness.   This allows us to address integer and fractional QHITs on equal footing.  As we describe, the effective field theoretic description that incorporates both disorder and interaction effects is a gauged non-linear sigma model (NLSM) with a non-zero topological term.  The NLSM  provides us with a framework to compute, at least in principle, fluctuational corrections to the conductivities in the neighborhood of the transitions.  As we show explicitly, the topological term, which governs the behavior of the Hall conductivity, behaves differently in the two CF theories, in the presence of gauge fluctuations.  In particular, in the Dirac CF theory, the topological term - and thus the Hall conductivity -  is unaffected by gauge fluctuations at criticality. Scaling behavior at criticality is therefore determined by two parameters: disorder ($1/\sigma^{\rm cf}_{xx}$) and gauge fluctuations, set by $e_*^2$, the strength of $1/r$ interactions (see Fig. \ref{pd}).

	While the critical point itself corresponds to a strong coupling fixed point of the NLSM,  we are nevertheless able to extract from it some robust properties of the critical point.  We  show that for the case of $1/r$ interactions, the transitions are {\it superuniversal}: both integer and fractional QHITs are in the same universality class.\cite{Engel1990, Kivelson1992, PhysRevLett.70.481, Shahar1995, SondhiGirvinCariniShahar}  Superuniversality was conjectured to hold due to a `law of corresponding states' relating integer and fractional quantum Hall phases\cite{Kivelson1992}.  The conditions underlying superuniversality had remained unknown. We provide an explicit discussion and show that superuniversality is guaranteed in the presence of $1/r$ interactions, provided the compressibility remains finite at the transition.  With short-ranged contact interactions, it is plausible that the transitions continue to exhibit superuniversality, but further analysis is needed to show  this unambiguously.

	Another important universal property that stems from interaction effects are dynamical scaling laws.  At a quantum critical point, a characteristic time scale $\xi_{\tau}$ diverges with the correlation length $\xi$ as $\xi_{\tau} \sim \xi^z$, where $z$ is the dynamical critical exponent.  The dynamical exponent plays an important role near quantum critical points, since it also affects the temperature dependence of thermodynamic quantities  ({\it e.g.} the singular contribution to the  heat capacity $C \sim T^{d/z}$), as well as conductivities.    In a strictly non-interacting description of QHITs, the only characteristic time scales comes from the behavior of the density of states (DOS) at the critical point: if the DOS remains finite, $z=d$ follows automatically.  With interactions, $z$ can change: for example, if $1/r$ interactions remain unscreened, then $z$ can be unity.  In the CF description, $z$ is determined by the behavior of transverse gauge fluctuations.  As we explicitly show, the gauge boson is overdamped both for $1/r$ interactions and short-ranged interactions.  But $z=1(z=2)$ for $1/r$(short-ranged) interactions.  The only requirement for this behavior is a finite conductivity at the critical point.  The finiteness of the dc longitudinal conductivity is guaranteed in our theory since gauge fluctuations do not renormalize the topological term (in Son's formulation of the CF theory), which in turn guarantees, by a corollary of Laughlin's gauge argument, that states at the Fermi energy are not localized.  
	
	The paper is organized as follows. We formulate the QHIT problem in the CF coordinates, including disorder and interactions in section \ref{sec:CF_theory}. A pedagogical derivation of the gauged non-linear sigma model description of QHITs is presented in section \ref{sec:NLSM}. We explain the origins of the dynamical scaling relation and superuniversality in sections \ref{sec:z=1} and \ref{sec:superuniversality} respectively. We summarize our findings and give concluding remarks in section \ref{sec:summary}. A derivation of the topological term in the Halperin-Lee-Read CF theory and the perturbative effects of gauge fluctuations are presented in the Appendices \ref{sec:HLR_top_term} and \ref{sec:perturbation_theory} respectively.

	\section{ QHITs in the composite fermion representation \label{sec:CF_theory}}
	The standard approach to the study of QH transitions involves electrons in a  perpendicular magnetic field with quenched disorder and  interactions:
	\begin{eqnarray}
	\mathcal L &=& \mathcal L_{0} + \mathcal L_{int}\nonumber \\
	\mathcal L_0 &=& c^{\dagger}(r) \left[ -i \partial_t + \mu + V(r) - \frac{1}{2m} \left( \bm \partial - i \bm A \right)^2 \right] c(r) \nonumber \\
	\mathcal L_{int} &=& -\frac{1}{2}\int d^2r' \left[ n(r) - \langle n \rangle \right] U(r-r') \left[ n(r') - \langle n \rangle \right].
	\end{eqnarray}
	The operator $c(r)$ destroys a spin-polarized electron at position $r$,  $n(r) = c^{\dagger} (r) c (r)$, $B = \nabla \times A$ is the perpendicular magnetic field, and the interaction potential is denoted by $U(r)$.  In addition, there is a quenched random potential $V(r)$, which couples to the Dirac electron density, and is obtained by  shifting $A_t \rightarrow A_t + V(\bm r)$. We assume that $V(\bm r)$ is zero on average and is statistically particle-hole symmetric, i.e. all odd moments of $V(\bm r)$ vanish.
While we focus mostly on $U(r) = e_*^2/r$, the case of experimental relevance, we also touch on the fate of $U(r) = U_0 \delta(r)$, {\it i.e.} screened Coulomb interactions.

The problem as formulated in electron coordinates is formidable; there are no small parameters, since one must contend with the large external magnetic field leading to Landau level quantization, and the effects of both interactions and disorder are crucial and singular.  Numerical approaches have suffered with the difficulty of going beyond the non-interacting limit, which, as we discussed above, is inadequate for describing many of the universal and experimentally relevant aspects of this problem.  

Our strategy then, is to consider a different coordinate system in which the combined physics of interactions and disorder are perhaps more tractable.  We therefore study the problem from the composite fermion (CF) perspective, which, as we show, provides new physical insights. 
	We use Son's Dirac CF formulation\cite{Son2015} of the half-filled Landau level, which we  briefly review in the next subsection.  We then show how QHITs are realized in the CF framework.  
		\subsection{Dirac CFs: brief review}

	For our purposes, it is sufficient to motivate the Dirac composite fermion theory by noting that the lowest Landau level (LLL) limit of a massless Dirac fermion is identical to that of a non-relativistic electron: despite the spinor nature of the Dirac wave functions, only one component is non-zero in the LLL and the dynamics are thus identical for Dirac and non-relativistic electrons.  Moreover, in the spirit of critical phenomena, universal aspects of quantum Hall transitions are insensitive to whether we choose to work with massless Dirac fermions or nonrelativistic particles in the LLL, each of which amount to different ultraviolet regulations of the same low energy physics.  Furthermore, as observed by Son, the massless Dirac electron  
	manifestly  preserves particle-hole (PH) symmetry without the need of the LLL limit. Additionally, one can explicitly impart a duality transformation tantamount to particle-vortex duality to obtain the Dirac CF theory:\cite{Seiberg:2016gmd}
	%
	\begin{gather}
		\mathcal{L}_{\rm Dirac\ el.} = i \bar c \gamma_{\nu} D_A^{\nu} c + \frac{1}{8\pi} AdA + \mathcal{L}_{\rm dis.} + \mathcal{L}_{\rm int.}\nn\\
		\Big\updownarrow\nn\\
		\begin{split}
			\mathcal{L}_{cf} &= \mathcal{ L}'_{\psi} + \mathcal{ L}'_{\rm gauge}+  \mathcal{L}'_{\rm dis.} + \mathcal{L}'_{\rm int.}
		\end{split}\label{eq:CF_lagrangian}
	\end{gather}\vspace{-5pt}
	\begin{align}
		\mathcal{ L}'_{\psi} &= i \bar \psi \gamma_{\nu} D_a^{\nu} \psi + \mu_{\rm cf} \bar\psi \gamma^t \psi\\
		\mathcal{ L}'_{\rm gauge} &= - \frac{1}{4 \pi} a d A + \frac{1}{8\pi} A d A\label{eq:lagr_gauge_CF}\\
		\mathcal{L}'_{\rm dis.} &= -\frac{1}{4\pi} V(\bm r)b\\
		\mathcal{L}'_{\rm int.} &= -\frac{1}{2(4\pi)^2} \int d^2r'\ U(|\bm r-\bm r'|)b(\bm r') b(\bm r) \label{eq:Son}
	\end{align}
	where $c,\bar c$ are the two-component Dirac electron operators, $\psi,\bar\psi$ are the corresponding Dirac composite-fermions operators, $D_a^\nu \equiv \partial^\nu - i  a^\nu $, $\gamma^\nu$ satisfy the relations $\{\gamma^\mu,\gamma^\nu\} = 2g^{\mu\nu}$ and $g^{\mu\nu}$ is the metric tensor with the signature $(-,+,+)$.  Further, $A$ represents the background electromagnetic gauge field while $a$ is the emergent gauge field.  The shorthand   $AdB = \epsilon_{\mu \nu \lambda} A_{\mu} \partial_{\nu} B_{\lambda}$ with greek indices $\mu = t,x,y$ and $\epsilon_{\mu \nu \lambda}$ is the Levi-Civita tensor.  
		The Chern-Simons term for $A$ in the Dirac electron Lagrangian density, reflects the  parity anomaly\cite{Redlichparitylong}: in a properly regularized theory, a single massless 2-component Dirac fermion in $d=2+1$ breaks time-reversal and has a half-integer Hall conductance in units of $e^2/h = 1/2\pi$.   Anticipating a finite density of CFs, we have introduced a CF chemical potential $\mu_{\rm cf}$.  The duality transformation above is stated in a continuum limit for simplicity, but it can formally be defined on a lattice and implemented as an exact mapping among lattice partition functions\cite{Son2019}.  We will, however, mainly be interested in the continuum formulation of the duality mapping as stated above. 
		\subsection{Disorder and interactions in the Dirac CF theory}
		
		Since $a_t$ appears linearly in the CF lagrangian, it acts as  a Lagrange multiplier and enforces the following constraint via its equation of motion:
	\begin{align}
		\rho_{\rm cf} \equiv \bar\psi \gamma^t \psi  = \frac{B}{4\pi} \label{eq:Son_cf_density}
	\end{align}
	Thus, the  CF density is fixed by the external magnetic field, and it is in this sense that the Dirac CFs are to be viewed as ``vortices" of the Dirac electron.

		Interactions in $\mathcal L_{int}$ are of density-density form; in the dual frame, they are captured in $\mathcal L_{int}'$, using the fact that the electromagnetic charge density is given by:
	\begin{gather}
		\rho_{\rm EM} \equiv \frac{\delta \mathcal L_{cf}}{\delta A_t} = \frac{B-b}{4\pi} \label{eq:Son_EM_charge}
	\end{gather}
	\begin{figure}
		\includegraphics[width=1.8in]{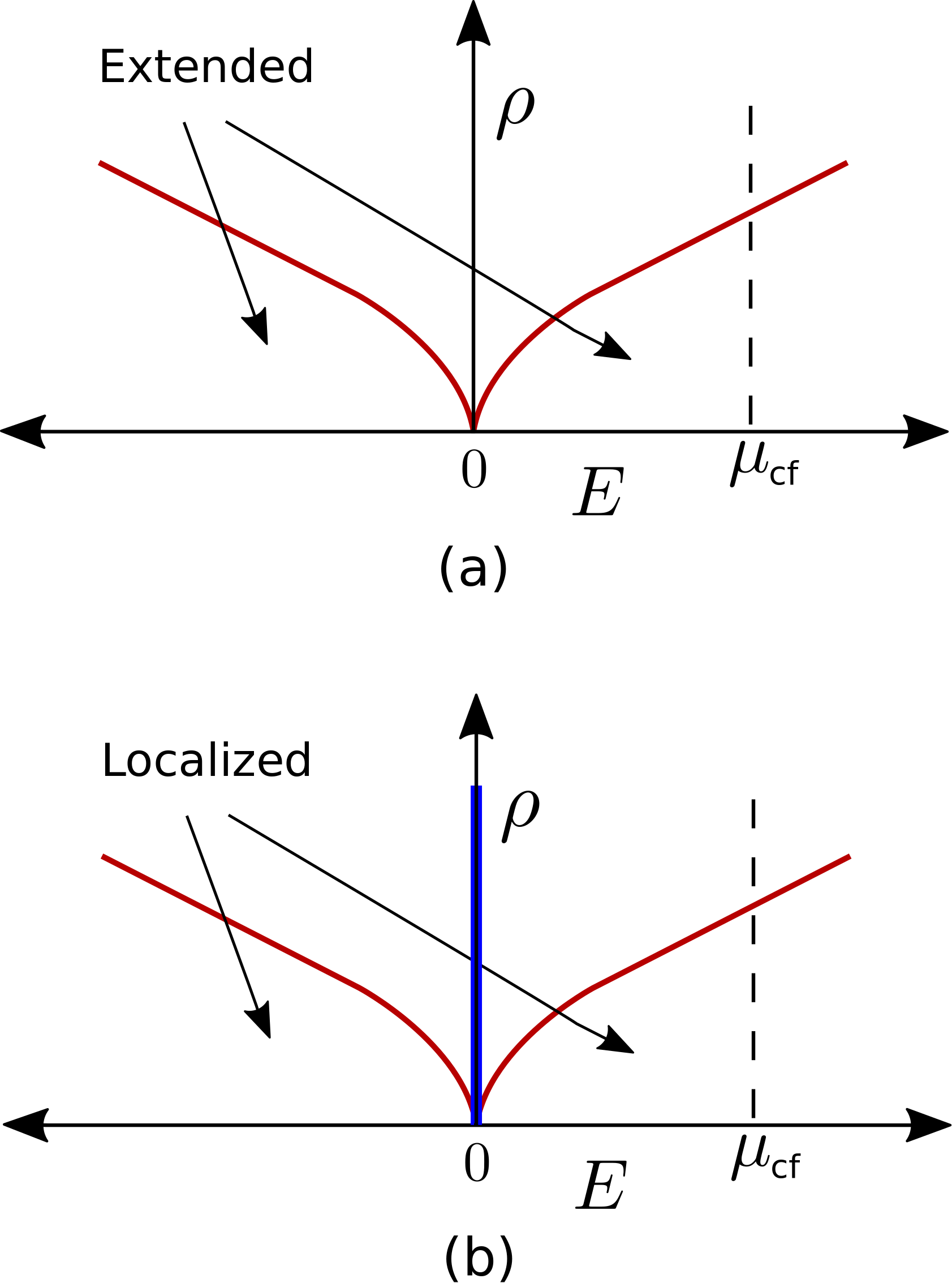}
		\caption{A schematic of the density of states of a Dirac fermion in the presence of a vector potential disorder $a'_j(\bm r)$. We define a tuning parameter $b_0 \equiv \frac{1}{L^2}\int d^2r\ \nabla \times \bm a'(\bm r)$. In subfigure (a), $b_0 = 0$, and the Hall conductivity is $\sigma_{xy}^{\rm cf} = 0$ at all Fermi-energies as a result of the average time-reversal symmetry. This implies that all states are extended due to Laughlin's gauge argument.\cite{Laughlin1981} In subfigure (b), $b_0 \neq 0$ and all the non-zero energy states are localized. There exist zero modes, the number of which is equal to total number of flux quanta passing through the system.\cite{Aharonov1979} These zero-modes lead to the Hall conductivity $\sigma_{xy}^{\rm cf} = \frac{\mathrm{sgn}[b_0]}{4\pi}$ at any positive Fermi-energy. Thus, the Dirac CFs exhibit an IQHT when we tune the average magnetic field across $b_0=0$.}
		\label{fig:IQHT_Son}
	\end{figure} 

	\subsection{Mean-field theory of the integer QHIT\label{sec:MFT_IQHIT}}
	
	Since the random potential $V(r)$ couples to the magnetic field $b$ in the dual theory, the $U(1)$ gauge field must contain a fluctuating piece, and a quenched random piece\cite{2017PhRvB..96x5140G}.  Thus, to accommodate quenched randomness in the CF representation, we shift  $a_{j}(r,t) \rightarrow   a_{j} (r,t) + a'_{j}(r)$, and $a'_{j}(r)$ are chosen from spatially independent gaussian random distribution,
	\begin{equation}
	\label{prob}
	P[a'] = e^{-\pi N_F \tau \int d^2 r a'(r)^2},
	\end{equation}
	where $N_F$ is the density of states at the Fermi level.  
	
	In the mean-field approximation, we neglect the fluctuations of the $U(1)$ gauge field beyond its equations of motion.  To formulate the mean-field theory carefully, it is useful to rearrange terms in part of the CF Lagrangian of Eq. \eqref{eq:CF_lagrangian} as:
	\begin{eqnarray}
	 i \bar \psi \slashed{D}_a \psi &+& \mu_{\rm cf} \bar\psi \gamma^t \psi- \frac{adA}{4 \pi} + \frac{AdA}{8 \pi}  \nn\\
	 &\,&\ \ \ \ \ \ \equiv \mathcal L_{\psi}[\bar\psi, \psi, a]+ \mathcal L_{gauge}[a,A], \nonumber \\
	\end{eqnarray}
with 
	\begin{eqnarray}
	\mathcal L_{\psi}[\bar \psi, \psi, a] &=& i \bar \psi \slashed{D}_a \psi+ \mu_{\rm cf} \bar\psi \gamma^t \psi - \frac{ada}{8 \pi} \nonumber \\
	\mathcal L_{gauge}[a,A] &=& \frac{(a-A)d(a-A)}{8 \pi},
	\end{eqnarray}
	so that the full CF theory takes the form
	\begin{equation}
	\mathcal L_{cf} = \mathcal L_{\psi}[\bar \psi, \psi, a]+\mathcal L_{gauge}[a,A]+\mathcal L_{dis}' + \mathcal L_{int}'.
	\end{equation}
	We do this in order to define a properly regularized Dirac theory, which can in principle be defined on a  lattice.  Indeed, the second term in $\mathcal L_{\psi}[\bar \psi, \psi, a]$ comes from the fact that on a lattice, there will be a massive doubler fermion, which generates a level 1/2 Chern-Simons term.  The electrical current $J_{\mu} = \delta \mathcal L_{cf}/\delta A_{\mu}$ is given  by
	\begin{equation}
	J_{\mu} = \frac{\epsilon_{\mu \nu \lambda}}{4 \pi} \partial_{\nu} \left(A_{\lambda} - a_{\lambda} \right),
	\end{equation}
	whereas the equation of motion of $a_{\mu}$ results  in
	\begin{equation}
	j^{\mu}_{\psi} \equiv  \frac{ \delta \mathcal L_{\psi}}{ \delta a_{\mu}}=\frac{\epsilon_{\mu \nu \lambda}}{4 \pi} \partial_{\nu} \left(A_{\lambda} - a_{\lambda} \right) 
	\end{equation}
	From these relations, and the definitions of linear response, $E_i = \rho_{ij} J_j,e_j = \rho^{\psi}_{ij} j_{\psi}^j$, it follows that the resistivity tensors satisfy
	\begin{equation}
	\label{dictionary1}
	\rho_{ab} = \rho_{ab}^{\psi}+ 4 \pi \varepsilon_{ab}, \ \ \varepsilon = \left( \begin{array}{cc} 0 & 1\\ -1& 0 \end{array} \right)
	\end{equation} 
	
	The single-particle  Hamiltonian associated with $L_{\psi}$ is also properly regularized, and involves  a massless Dirac fermion along with a massive doubler fermion:
		\begin{eqnarray}	
		\mathcal H_{\rm cf} &=& \mathcal H_1 + \mathcal H_2 \nonumber \\
		 \mathcal H_1 &=& \bm \sigma.(\bm p - \bm a') - \mu_{\rm cf} \nonumber \\
		 \mathcal H_2 &=& \bm \sigma.(\bm p - \bm a') - m \sigma^z - \mu_{\rm cf}, \ \ m \gg \mu_{\rm cf}
	\end{eqnarray}
	where $\bm a' \equiv \bm a'(\bm r)$ is again the quenched random vector potential, the Dirac fermions are at a finite chemical potential $\mu_{\rm cf}$ in accordance to Eq. \eqref{eq:Son_cf_density} and $\bm\sigma = {\sigma_x,\sigma_y}$ are the Pauli matrices.  Having defined the theory properly, we may now safely ignore $\mathcal H_2$ since the mass term is large compared to the chemical potential.  It will play a role in contributing $-1/4 \pi$ to the CF Hall conductivity.

	The mean-field phase diagram is then obtained by permitting the  quenched random magnetic field to have a non-zero  time-independent average  $b_0$:
	\begin{equation}
	b_0 \equiv \frac{1}{L^2} \int d^2r\ \nabla\times \bm a'(\bm r).
	\end{equation} 
	This quantity acts as the tuning parameter for the QHIT, and the critical point occurs at $b_0 = 0$.  
	
	An essential role is played by the Dirac CF {\it zero modes}\cite{Aharonov1979} (Fig. \ref{fig:IQHT_Son}).  When $b_0 \neq 0$, the  zero modes of $\mathcal H_1$ are analogous to the lowest Landau level\cite{Kumarsusy} and contribute a Hall conductance of $\frac{\mathrm{sgn}[b_0]}{4\pi}$. Due to disorder,the non-zero energy states are localized when $b_0 \ne 0$. Therefore, the total Hall conductance of the CFs, including the contribution from the massive doubler in  $\mathcal H_2$ is given by:
	\begin{align}
		\sigma_{xy}^{\rm cf} = \left\{\begin{matrix}
			\frac{\mathrm{sgn}[b_0]-1}{4\pi}, & b_0 \neq 0\\
			-\frac{1}{4 \pi} & b_0 = 0
		\end{matrix}\right.
	\end{align}
	which when translated to electrical response via Eq. \ref{dictionary1}, corresponds to the $\nu = 1\rightarrow 0$ transition.  This is the CF formulation of the integer QHIT, within mean-field theory.

	A crucial property of the  CF mean-field theory is that at criticality, {\it delocalized states occur at all energies}: states at the Fermi level remain compressible and give rise to a non-zero $T=0$ dc conductivity.  This will play a crucial role in the conclusions that follow.  The existence of  delocalized states follows from the parity anomaly of the  Dirac fermion: a half-integer Hall conductivity arising from UV degrees of freedom (on a lattice, it originates from massive doubler fermions) ensures, by a corollary of Laughlin's gauge argument\cite{Laughlin1981} that states must be delocalized, irrespective of the Fermi level.  As we show in section \ref{sec:NLSM}, this feature continues to hold with gauge fluctuations.  

	An important universal property of the QHIT are the critical exponents associated with diverging length and time scales.  Since all states at the chemical potential are localized for $b_0 \ne 0$, the characteristic length scale, the localization length $\xi$, diverges as $\vert b_0 \vert^{-\nu}$, where $\nu$ is  a critical exponent (not to be confused with the filling fraction).  The non-interacting QHIT has been studied extensively in the electron representation via numerical simulations of the Chalker-Coddington model.   
	In a recent numerical study of the mean-field IQHIT in the dual CF representation, 2 of us have found $\nu = 2.56\pm 0.02$ in agreement with numerical studies in the electron representation.\cite{Huang2021}  Thus, this provides us with confidence that the CF mean-field theory above faithfully captures the relevant physics.   Finally, as mentioned in the introduction, the dynamical critical exponent governing the critical slowing down is determined solely by the finite density of states at the chemical potential, and $z = d = 2$ for the CF mean-field theory as well as the electron counterpart.

We next discuss the mean-field description of abelian fractional QHITs. 

\subsection{Abelian fractional QHITs}
	A distinctive advantage of the CF approach over the electron approach\cite{Liu2016} is that it enables us to study fractional and integer QH transitions on equal footing.  We will consider here  transitions from a class of abelian fractional QH states with filling fraction
	$\nu = 1/(2m -1), m=1,2, \cdots$, to insulating phases.  In the CF representation, the effective theory takes the same form, namely, 
	\begin{equation}
	\mathcal L_{cf} = \mathcal L_{\psi}[\bar \psi, \psi, a]+\mathcal L_{gauge}[a,A]+\mathcal L_{dis}' + \mathcal L_{int}',
	\end{equation}
	where $\mathcal L'_{dis}, \mathcal L'_{int}$ and $\mathcal L_{\psi}$ are the same as before, but $\mathcal L_{gauge}$ is generalized to\cite{Goldman2018}
	\begin{equation}
	\label{lgauge}
	\mathcal L_{\rm gauge} = \frac{1}{4 \pi} \frac{1}{2m} \left(a - A \right) d \left( a - A \right)
	\end{equation}
	For $m=1$, we recover the description of the integer QH transition.  
	
	The mean-field description of  fractional QHITs proceeds identically as before.   We include the contribution to $a_j$ from quenched disorder and treat all gauge fluctuations at the level of the classical equations of motion.    A consequence of this approximation is that the behavior from $\mathcal L_{\psi}$ is the same as in the IQHIT, with the only change being the relation between the resistivity tensors, due to the change in $\mathcal L_{gauge}$:
	\begin{equation}
	\rho_{ab} = \rho_{\psi, ab}+ 4 \pi m \varepsilon_{ab}.
	\end{equation}
	This condition, in conjunction with the fact that the Dirac fermions undergo a $\nu = -1 \rightarrow 0$ transition, captures the correct mean-field phase diagram.

		The fact that in the mean-field approximation,  the sector describing the Dirac CFs is unchanged in both the IQHIT and the FQHIT case has significance.  It implies that the localization length exponent $\nu$ and the dynamical exponent $z$ is trivially the same for both cases.  This is the sense of `superuniversality' that was put forward in Ref. \onlinecite{Kivelson1992}.  However, at the level of  CF mean-field theory it is not surprising - since the behavior of CFs  in the disordered environment is identical in both cases.  Such superuniversality is somewhat reminiscent of Landau's mean-field theory of phase transitions, in which all continuous transitions have the same mean-field exponents regardless of their nature.  A far less trivial issue is the extent to which interaction effects preserve  superuniversality.  This has proven to be notoriously difficult to address in electron coordinates.  As we show, the CF formulation readily addresses this issue.

\subsection{QHITs with Gauge fluctuations}	
	
	Next, we take into account the interaction lagrangian density, $\mathcal L'_{int}$, which in CF  coordinates restores gauge fluctuations, since $\mathcal L'_{int}$  sets the kinetic term for the transverse gauge boson:
	\begin{equation}
	\mathcal L'_{\rm int} = -\frac{1}{2}\int \frac{d^2 r '}{\left( 4 \pi \right)^2} b(r,t) U(x-x') b(r',t).
	\end{equation}
	In Coulomb gauge, the dynamical fields have a temporal $a_0$ and transverse components $a_T$.
	Interactions $U(r)$ depend only on the transverse gauge boson: for $U(r) = e_*^2/r$, the interaction lagrangian takes the form
	\begin{gather}
	\label{Lint}
	L'_{\rm int} = -\frac{e_*^2}{16\pi} \int d^2q\ \vert q \vert  a_T(-q)   a_T (q)\\
	a_j( q) = \frac{i\epsilon^{jk}q_j a_T( q)}{q}
	\end{gather}
	For the case of contact interactions, $U(r) = U_0 \delta(r)$, 
	\begin{equation}
	L'_{\rm int} = -\frac{U_0}{16\pi} \int d^2 q\  q^2   a_T(-q)   a_T (q)
	\end{equation}
	With short-range interactions, gauge fluctuation effects are stronger and the interplay between disorder and interactions remains more subtle.  We will discuss the effects of short-ranged interaction on QHITs in section \ref{sec:superuniversality}.

\section{Composite fermion nonlinear sigma model \label{sec:NLSM}}
		
	At length scales large compared to the mean-free path, the CFs are strongly scattered by the disorder and the diffusive modes are the emergent low energy degrees of freedom. The nonlinear sigma model (NLSM)\cite{Belitz1994, Nayaklectures, kamenev2011field, Pruisken1999} serves as a framework to study the interplay of these diffusive modes and the gauge fluctuations. In this section, we present a pedagogical derivation of the NLSM in the CF coordinates that will form the basis of further theoretical analysis of the Abelian FQHITs.

	We work in imaginary time and disorder average using the replica trick. Ignoring the external fields for the moment, the partition function of the replicated Son's theory is:
	\begin{align}
	\mathcal{Z}^{N_r} &= \int \prod_{\alpha=1}^{N_r} \mathcal{D}[\bar\psi^\alpha,\psi^\alpha]\mathcal{D}[a_\mu^\alpha]\ e^{-S}\nn\\
	S &= S^\psi + S^{\rm dis.} + S^{a}\nn\\
	S^\psi &= -\sum_{\alpha,n} \int d^2r\ {\psi^\dagger}_n^\alpha\left(i\omega_n  + \mu  +i\sigma_j\partial_j \right)\psi_n^\alpha\nn\\
	&\ \ \ \ \ \ \ -\frac{i}{\sqrt{\beta}}\sum_{n,m,\alpha}\int d^2r\ a_\mu^\alpha(\omega_n-\omega_m){\psi^\dagger}_n^\alpha \gamma^\tau\gamma^\mu \psi_m^\alpha  \nn\\
	S^{\rm dis.} &=  - \sum_{\alpha,n}\int d^2r\ a'_j(\bm r){\psi^\dagger}^\alpha_n \sigma_j \psi^\alpha_n\nn \\
	S^{a} &= \frac{1}{2(4\pi)^2} \sum_\alpha\int d^2r\ d^2r'\ U(|\bm r-\bm r'|)b^\alpha(\bm r') b^\alpha(\bm r) 
	\end{align}
	where $N_r$ is the number of replicas, $\psi^\dagger \equiv -\bar\psi \gamma^\tau$, $\gamma^\tau = \sigma_z, \gamma^\tau\gamma^j = -i\sigma_j$, $\omega_n = \frac{(2n+1)\pi}{\beta}$ and $\beta$ is the inverse temperature. The greek letter superscripts correspond to the replica indices and $n,m$ are the frequency indices. Further, the Fourier transform convention is:
		\begin{gather}
			\psi_n(\bm q)  \equiv \frac{1}{\sqrt\beta} \int d^2r \int_0^\beta d\tau\ e^{i\omega_n\tau - i\bm q.\bm r} \psi(\bm r, \tau)\label{eq:Fourier_convention}
		\end{gather}
	
	Using the probability distribution in Eq. \eqref{prob} to perform disorder averaging, we obtain the following four-fermion interaction:
	\begin{align}
		S^{\rm dis.} &=  -\frac{1}{2\pi N_F \tau}\int d^2r \left(\sum_{\alpha,n}\ {\psi^\dagger}^\alpha_n \sigma_j \psi^\alpha_n\right)^2\nn\\
		&= 	\frac{1}{2\pi N_F \tau} \sum_{nm,\alpha\beta} \int d^2r \ \left[({\psi^\dagger}^\alpha_n\psi^\beta_m) ({\psi^\dagger}^\beta_m \psi^\alpha_n)\right.\nn \\
		&\ \ \ \ \ \ \ \ \ \ \ \ \ \ \ \ \ \left.-  ({\psi^\dagger}^\alpha_n\sigma_z\psi^\beta_m) ({\psi^\dagger}^\beta_m\sigma_z \psi^\alpha_n)\right]
	\end{align}
	The second term is unimportant since the Dirac spinor magnetization is not conserved. We decouple the first term using a Hubbard-Stratonovic field: $Q = Q^\dagger \equiv Q^{\alpha\beta}_{nm}(\bm r)$ and obtain the following action:
	\begin{align}
		S &= S^{\psi,Q} + S^a \nn\\
		S^{\psi,Q} &= \int d^2r\ \left[-\sum_{nm,\alpha\beta}{\psi^\dagger}_n^\alpha \left[ {(G_0^{-1})}^{\alpha\beta}_{nm} + \frac{i}{2\tau}Q^{\alpha\beta}_{nm}\right.\right.\nn\\
		&\ \ \ \left.\left.  +\frac{1}{\sqrt{\beta}}\left(i \mathcal{A}^{\alpha\beta}_{\tau,nm} + \sigma_j \mathcal{A}^{\alpha\beta}_{j,nm}\right)  \right] {\psi}_m^\beta  + \frac{\pi N_F}{4\tau} \mathrm{Tr}[Q^2]\right] \label{eq:gauged_disorder_avg}
	\end{align}\vspace{-20pt}
	\begin{align}
	G_0^{-1} &\equiv i\Omega + \mu +i\sigma_j\partial_j\\
	\mathcal{A}_{\mu,nm}^{\alpha\beta} &\equiv  a_{\mu}^\alpha(\omega_n-\omega_m) \delta^{\alpha\beta}\\
	\Omega^{\alpha\beta}_{nm} &\equiv \omega_n \delta_{nm}\delta^{\alpha\beta}
	\end{align}
	
	$Q$ has a non-zero saddle point value $Q=\Lambda$ that gives rise to a finite lifetime for the CFs. In addition, the Goldstone mode fluctuations around this saddle point give rise to the diffusive degrees of freedom. As a first step in deriving the effective field theory of QHIT, we use the equation of motion of $Q$ to obtain the saddle point solution:
	\begin{align}
		Q &= \frac{i}{\pi N_F}\int d^2r\ \mathrm{Tr}_{\sigma}\left[\frac{1}{G_0^{-1} + \frac{i}{2\tau}Q}\right]\\
		\Lambda &= \mathrm{sgn}[\Omega]
	\end{align}
	where $\mathrm{Tr}_{\sigma}$ indicates trace over the spinor indices.
	
	The Goldstone modes result from the fact that to zeroth order in $\mathcal{A}$ and $\Omega$, the action in Eq. \eqref{eq:gauged_disorder_avg} is invariant under the unitary transformation $U \equiv U^{\alpha\beta}_{nm}$: $\psi \rightarrow U\psi, \psi^\dagger \rightarrow \psi^\dagger U^\dagger, Q\rightarrow UQU^\dagger$. And, the saddle-point solution $Q= \Lambda$ spontaneously breaks the $U(2N_r N_\omega)$ symmetry down to $U(N_r N_\omega)\times U(N_rN_\omega)$. Where we consider a frequency cutoff $\Omega_c$ such that $|\omega_n| < \Omega_{c}$ and $2N_\omega$ is the number of Matsubara frequencies inside this interval. 
	The unitary symmetry implies that the full set of saddle-point solutions, i.e. $U\Lambda U^\dagger$, is given by the following conditions:
	\begin{gather}
		Q^2 = \mathbb{1}, \ \ \ \ \ \ \ 
		\mathrm{Tr}[Q] = 0
	\end{gather}

	We now obtain the long wavelength, low energy effective field theory by expanding the action to leading order in the variations of $Q$ and the gauge field. As mentioned earlier, it is a NLSM of the diffusive modes interacting with a gauge field. Let's first integrate out the fermions:
	\begin{align}
	S^{\psi,Q} &\rightarrow  - \mathrm{Tr}\log\left[G_0^{-1}+ \frac{i}{2\tau}Q+\frac{1}{\sqrt{\beta}}\left(i \mathcal{A}_{\tau} + \sigma_j \mathcal{A}_{j}\right)\right] \nn\\
	&\ \ \ \ \ \ \ \ \ +  \frac{\pi N_F}{4\tau}\int d^2r \ \mathrm{Tr}[Q^2]
	\end{align}

	There are a few critical issues that we need to account for. First, the effective action obtained by derivative expansion must preserve gauge invariance.\cite{kamenev2011field} To this end, we note that the low energy degree of freedom transforms as $Q_{tt'} \rightarrow e^{i\Gamma(\bm r, t)}Q_{tt'}e^{-i\Gamma(\bm r, t')}$ under the gauge transformation: $\psi \rightarrow e^{i\Gamma(\bm r, t)}\psi,\ a_\mu \rightarrow a_\mu + \partial_\mu \Gamma$.
	
	Second, Son's theory contains a single Dirac fermion. We interpret the frequency and replica indices as flavor indices and thus the Dirac fermion is $2+0$-dimensional. As such, it contains a chiral anomaly that gives rise to an imaginary term in the action. In subsection \ref{sec:top_term_derivation}, we show that this is, in fact, the topological term proposed by Pruisken\cite{Pruisken1984} with a coefficient that corresponds to the QHIT.  We provide an explicit and pedagogical derivation of the topological term of the NLSM.  
	
	With these considerations, we obtain the gauged NLSM action:
	\begin{align}
		S &= S^{Q} + S^{\rm top} + S^a \label{eq:gauged_NLSM}\\
		S^{Q} &= \pi N_F\int d^2r\ \mathrm{Tr}\left[\frac{ D}{4}(D_j Q)^2 - \left(\Omega+\frac{\mathcal{A}_\tau}{\sqrt{\beta}}\right) Q \right]\\
		S^{\rm top} &= \frac{1}{16} \int d^2r\ \epsilon^{jk}\mathrm{Tr}[Q\partial_jQ \partial_k Q]\\
		S^a &=  \frac{1}{2(4\pi)^2} \sum_\alpha\int d\tau d^2rd^2r'\ U(|\bm r-\bm r'|)b^\alpha(\bm r') b^\alpha(\bm r)\nn\\
		&\ \ \ \ \ \ \ \ \ \   + \frac{N_F}{2} \sum_\alpha\int d^2rd\tau\ {(a_\tau^\alpha)}^2 \label{eq:saddle_point_a}\\
		D_jQ &\equiv \partial_j Q - \frac{i}{\sqrt{\beta}}[\mathcal{A}_j, Q]
	\end{align}
	where $D \equiv v^2\tau/2$ is the diffusion constant and $v=1$ is the Fermi velocity. The second term in Eq. \eqref{eq:saddle_point_a} results from the Debye screening due to the finite compressibility of the CFs.
	
	As alluded to earlier, the coefficient of the topological term, $S_{\rm top}$, corresponds to the CF theory being at the IQHIT from $\nu_{\rm cf} = -1$ to $\nu_{\rm cf} = 0$. This confirms our expectations of section \ref{sec:MFT_IQHIT} in the NLSM language. Remarkably, for Dirac CFs, the topological term is unaffected by the gauge field (see section \ref{sec:top_term_derivation} and Ref. \onlinecite{Mirlin2014}).  It reflects the fact that at criticality, the massless component of the Dirac fermion has ``time-reversal symmetry"\footnote{By time-reversal symmetry, we mean that the contribution from the light Dirac fermions to the Hall conductance is identically zero.  For further details, see Ref. \onlinecite{Son2015}}, and any Hall conductance only arises from the massive UV degrees of freedom, which cannot couple to low energy diffusive fluctuations.  This  property of Dirac fermions guarantees that   {\it gauge fluctuations do not renormalize the Hall conductivity at criticality}.  Thus, by Laughlin's gauge argument, we are guaranteed of the existence of delocalized states at the Fermi level, even as the theory itself flows to strong coupling.  We note in passing, that in  other CF theories, such as that of Halperin, Lee and Read,\cite{Halperin1993} the contribution to $\sigma^{\rm cf}_{xy}$ is entirely from low energy quasiparticles, which {\it do} couple to diffusive modes.  In this case the CF Hall conductance {\it is} renormalized by gauge fluctuations, as we show in appendix \ref{sec:HLR_top_term}.
	
	Solving the NLSM for the realistic case of finite interactions strength and $\sigma_{xx}^{\rm cf} \sim \mathcal{ O}(e^2/h )$ is prohibitively difficult. Nevertheless, we can study the effect of gauge fluctuations on the diffusive degrees of freedom in the limit $1/\sigma_{xx}^{\rm cf}\rightarrow 0, \ 1/{e_*}^2 \rightarrow 0$ using perturbation theory where $e_*^2$ corresponds to the strength of interactions. In appendix \ref{sec:perturbation_theory}, we compute the corrections to the CF longitudinal conductivity due to the gauge fluctuations arising from Coulomb interactions in this limit and show that they are irrelevant. Therefore, the RG flow in this regime is towards the non-interacting CF limit and suggests the schematic RG flow shown in Fig. \ref{pd}. 
	
	An important distinction between the traditional electron version of IQHIT and the CF theory presented in this paper is that the perturbation theory in Coulomb interaction is controlled by the parameters $e^2_*$ and $1/e^2_*$ respectively. The two theories, therefore, correspond to two different limits of the interacting IQHIT problem. The $1/e^2_* \rightarrow 0$ limit, where the CF theory is applicable, has the advantage of being able to treat integer and fractional QHITs at equal footing. As we show in the following sections, it is a better starting point since it can naturally explain the $z=1$ dynamical scaling and superuniversality of QHITs.

	We considered a pure vector potential disorder in this section. However, in addition, the Dirac fermions may experience a random mass\cite{Mitra2019} and random chemical potential. For this generalized disorder, the general form of the effective field theory is the same as the one derived in this section because the composite-fermion density is the only slow mode and the NLSM remains in the unitary symmetry class.

	\subsection{Topological term in the Dirac CF theory\label{sec:top_term_derivation}}
	In this subsection, we show that the chiral anomaly of Dirac fermion contributes an imaginary term in the NLSM action that is identical to the topological term. Following Ref. \onlinecite{Kumar2019b}, we start with the following term in the action of Eq. \eqref{eq:gauged_disorder_avg}:
	\begin{align}
	S^\psi &= -\int d^2r\ {\psi^\dagger} \left[ i\Omega + \frac{i\mathcal{ A}_\tau}{\sqrt{\beta}}+ \mu \right.\nn\\
	&\ \ \ \ \ \ \ \ \ \ \ \ \left.+ i\sigma_j\partial_j + \frac{i}{2\tau}Q+\frac{\sigma_j \mathcal{A}_{j}}{\sqrt{\beta}}  \right] \psi
	\end{align}
In what follows, we'll take $\Omega \rightarrow 0$ and $\mathcal{A}_\tau \rightarrow 0$ for a leading order analysis. Thus, the Dirac fermion essentially becomes $2+0$-dimensional with the temporal and replica components playing the role of flavor indices. Upon parameterizing the saddle point $Q = u\Lambda u^\dagger$ and transforming fermions via a unitary transformation $\psi \rightarrow u\psi, \psi^\dagger \rightarrow \psi^\dagger u^\dagger$, we obtain:
	\begin{align} 
	S^\psi &= -\int d^2r\ {\bar\psi} \left[  \mu + i\sigma_j\partial_j  + \frac{i}{2\tau}\Lambda + \sigma_j B_j \right] \psi \\
	B_j &\equiv iu^\dagger\left(\partial_j - i \frac{\mathcal{A}_j}{\sqrt{\beta}}\right)u
	\label{Bj}
	\end{align}
	where\footnote{In the expression for $B_j$ in Eq. \ref{Bj}, derivatives are not operators; they only act directly on $u$.} $\psi^\dagger$ has been reinterpreted as $\bar\psi$ in the $2+0$-dimensional theory. We partition the gauge field $B_j$ into two components:
	\vspace{-5pt}
	\begin{align}
	B_j &= C_j + H_j\\
	C_j &\equiv \frac{1}{2} \left(B_j + \Lambda B_j \Lambda\right)\\
	H_j &\equiv \frac{1}{2} \left(B_j - \Lambda B_j \Lambda\right)\\
	[C_j, \Lambda] &= \{H_j,\Lambda\} = 0\\
	D_jQ &= \partial_j Q - \frac{i}{\sqrt\beta}[\mathcal{A}_j,Q] = iu[\Lambda, B_j]u^\dagger \label{eq:Goldstone_gauge_parts}
	\end{align}
	Gauge transformations that commute with $\Lambda$ don't modify the saddle point, this implies that $C_j$ are unbroken gauge degrees of freedoms while $H_j$ are Goldstone modes as seen from Eq. \eqref{eq:Goldstone_gauge_parts}. So, we can drop $H_j$ in the chiral anomaly analysis. 
	
	The mass matrix for the $2+0$-dimensional Dirac fermions is $M = \Lambda/2\tau -i\mu$. The masses for positive frequencies and negative frequencies are related as $m_+ = -m_-^*$. We make these masses complex conjugates of each other via the chiral rotation $\psi \rightarrow U(\alpha)\psi, \bar\psi \rightarrow \bar\psi U(\alpha)$, where $\alpha = 1$, $U(\alpha) \equiv e^{-i\alpha\gamma^5(\Lambda - \mathbb{1})\frac{\pi}{4}}$ and  $\gamma^5 \equiv i\sigma_x\sigma_y = -\sigma_z$. The action transforms to:
	\begin{gather}
	S^\psi = -\int d^2r\ {\bar\psi} \left[  \mu\Lambda +\frac{i}{2\tau} + i\sigma_j\partial_j  + \sigma_j C_j \right] \psi
	\end{gather}
	The Jacobian $J$ of the transformation leads to an imaginary piece in the action\cite{Fujikawa1979}:
	\begin{align}
	J &= \exp \left[-\frac{i}{2\pi }\frac{\pi}{4}\int d^2r \ \mathrm{Tr}\left[\gamma^5(\Lambda-\mathbb{1}) \slashed D^2\right]\right]\nn\\
	&= \exp \left[\epsilon^{jk}\frac{i}{2\pi }\frac{\pi}{4}\int d^2r \ \mathrm{Tr}\left[(\Lambda-\mathbb{1}) F^C_{jk}\right]\right]\\
	F^C_{jk} &\equiv \partial_j C_k - \partial_k C_j - i[C_j,C_k]
	\end{align}
	After some straightforward algebra, we obtain:
	\begin{gather}
	F^C_{jk} = \frac{1}{2\sqrt{\beta}}\left[u^\dagger F^\mathcal{A}_{jk}u + \Lambda u^\dagger F^\mathcal{A}_{jk}u\Lambda\right] -\frac{i}{4} \left[ \left[B_j, \Lambda  \right], \left[B_k, \Lambda \right]  \right]\\
	F^\mathcal{A}_{jk} \equiv \partial_j\mathcal{A}_k - \partial_k\mathcal{A}_j.
	\end{gather}
	We next employ the idea of {\it anomaly matching}, namely that the chiral anomaly induced jacobian is present at all length scales and does not undergro renormalization group flow.  Consequently, it can be expressed in terms of either the UV or the IR degrees of freedom (the latter in our case correspond to the matrix $Q$).  More explicitly, using Eq. \ref{eq:Goldstone_gauge_parts}, we can express the field strength as
	\begin{align}
	F^C_{jk} = \frac{1}{2\sqrt{\beta}}\left[u^\dagger F^\mathcal{A}_{jk}u + \Lambda u^\dagger F^\mathcal{A}_{jk}u\Lambda\right] +\frac{i}{4} u^{\dagger} \left[ D_j Q, D_k Q \right] u.
	\end{align}
		We then  obtain $J = e^{-S_{\rm top}}$ with
	\begin{align}
	S_{\rm top} &= \frac{\epsilon^{jk}}{16}\int d^2r \ \left[\mathrm{Tr}\left[QD_j Q D_k Q\right] - \frac{2i}{\sqrt{\beta}}\mathrm{Tr}[Q F^\mathcal{A}_{jk} - F^\mathcal{A}_{jk}]\right]  \nn\\
	&= \frac{\epsilon^{jk}}{16}\int d^2r \ \mathrm{Tr}\left[Q\partial_j Q \partial_k Q\right]\label{eq:top_term_Son} 
	\end{align}
	We have dropped the $\mathrm{Tr}[F^\mathcal{A}_{jk}]$ term since it is a constant. 
	
	To sum up, we have explicitly derived the topological term of the NLSM using anomaly matching, and conclude that the topological term consists of {\it ordinary} derivatives, not covariant derivatives.  This implies that at criticality, gauge fluctuations do not affect the running of $\theta$.  In the appendix, we provide a similar explicit derivation of the topological term for the composite fermion theory of Halperin, Lee and Read.  Interestingly, we find that the topological term in that theory is expressed in terms of {\it covariant} derivatives.  Thus, gauge fluctuations {\it will} affect the Hall conductivity at criticality in HLR theory.  
	
	\section{Dynamical scaling law\label{sec:z=1}}

	Having explicitly derived the NLSM in the composite fermion representation in the previous section, we deduce some of its robust consequences in the remainder of the paper.  We first consider the leading effects of gauge fluctuations in the presence of interactions and disorder.  
	The first important quantum corrections involve Debye screening of $a_\tau$, due to the finite compressibility of the critical CF metal, and overdamping of the transverse gauge boson.  In the clean limit, Landau damping sets the dynamical scaling laws.  By contrast, with disorder, and by the existence of delocalized states at criticality, the Kubo formula implies that the quantum correction to the transverse gauge boson via the current-current correlator is simply $\Pi^{TT}(q, \omega) = |\omega| \sigma_{xx}^{\rm cf}$, where $\sigma_{xx}^{\rm cf}$ is the conductivity of CFs.  As a consequence, the transverse gauge boson is overdamped with $z=1$ scaling with inverse propagator
	\begin{equation}
	D_T^{-1}(q,\omega) =  |\omega| \sigma_{xx}^{\rm cf}+ \frac{e_*^2}{8\pi} \vert q \vert.
	\end{equation}
	We can understand the $z=1$ scaling from a complementary perspective of overdamped plasmons as follows.\cite{Chow1996}  At criticality, the non-zero CF conductivity in turn implies a non-zero electrical conductivity $\sigma_{xx} = 1/(4\pi)^2\sigma_{xx}^{\rm cf}$.  It follows from this, that Coulomb interactions alter the relaxation of the electric charge density  $\rho_e$ via
	\begin{align}
	i \omega_R \rho_e &= 2\pi\sigma_{xx} q^2 U(\omega_R, q) \nn\\
	&= 2\pi e_*^2\sigma_{xx} |q| \rho_e + \cdots \label{eq:plasmon_dispersion}\\
	&= \frac{e_*^2}{8\pi \sigma_{xx}^{\rm cf}} |q|\rho_e + \cdots
	\end{align}
	where $\omega_R$ refers to the frequency in the real time. The first equation can be derived as follows. An electric charge density fluctuation creates a spatially varying Coulomb potential $U(\omega_R, q) = e_*^2 \rho(\omega_R, q)/|q|$ and the system responds to this electric field by generating the relaxing current: $J^i = \sigma_{ij} (-iq_j U)
	$. This in conjunction with the continuity equation, i.e. $i\omega_R \rho_e = iq_i J^i$ gives the plasmon dispersion of Eq. \eqref{eq:plasmon_dispersion}. \footnote{Such plasmon type reasoning was also proposed to be the source of $z=1$ scaling for superconductor to insulator transition in Ref. \onlinecite{Fisher1990}.}  
	The dynamics of plasmons are overdamped as opposed to reactive, i.e., they are dissipative. As such, we note that the $z=1$ scaling relation is not equivalent to Lorentz invariance.

	An interesting prediction of the plasmons being the source of $z=1$ scaling is that the overdamped dispersion is independent of the disorder strength, assuming $\sigma_{xx}^{\rm cf}$ flows to a universal value at the QHIT fixed point. This may explain the observations of Ref. \onlinecite{Li2009}, where the data taken from different samples collapse on the $z=1$ scaling curve, despite the fact that the individual samples contain varying amounts of disorder. \cite{Pruisken2009comment}

	Although the transverse gauge propagator can be obtained to leading order in the non-linear sigma model, we emphasize that the $z=1$ scaling relation only requires a non-zero  linear response conductivity of CFs. It is thus a robust property of the critical point that occurs at $\sigma_{xx}^{\rm cf} = \mathcal{ O} (e^2/h)$.
	
	\section{ Abelian fractional QH transitions: superuniversality\label{sec:superuniversality}}

	A distinctive advantage of the CF approach over the electron approach\cite{Liu2016} is that it enables us to study fractional and integer QH transitions on equal footing.  We will consider here  transitions from a class of abelian fractional QH states with filling fraction
	$\nu = 1/(2m -1), m=1,2, \cdots$, to insulting phases.  In the CF representation, we generalize $\mathcal L'_{\rm gauge}$ of Eq. \ref{eq:lagr_gauge_CF} to\cite{Goldman2018}
	\begin{equation}
	\label{lgauge}
	\mathcal L'_{\rm gauge} = -\frac{i}{4 \pi}\left( \frac{1-m}{2m} \right)ada + \frac{i}{2 \pi} \frac{1}{2m} a d A - \frac{i}{4 \pi} \frac{1}{2m} AdA.
	\end{equation}
	where we have Wick rotated to Euclidean time. For $m=1$, we recover the description of the integer QH transition; for $m>1$ there is now  a Chern-Simons (CS) term for $a_{\mu}$.  
	
	With $1/r$ interactions,  all transitions are in the same universality class regardless of the value of the integer $m$.  The reason for this, is that at criticality, the CF metal remains  {\it compressible} even with quenched randomness.  As a result, the CF degrees of freedom exhibit Debye screening.  More explicitly, the temporal component $a_\tau$ obtains a quantum correction $\Pi^{\tau\tau}(q, \omega) = N_F + \cdots$, and fluctuates above a ``massive" saddle, where $N_F$ is the density of states at the Fermi energy.  In Coulomb gauge, the CS term takes the form $\lambda a_\tau q a_T$, where $q$ is the transverse momentum, and $\lambda$ is given by the first term in Eq. \eqref{lgauge}.   Upon integrating out $a_\tau$, the effective Lagrangian for $a_T$ (leaving aside their coupling to CF degrees of freedom for the moment) takes the form
	\begin{align}
	\mathcal L'_{\rm eff}[a_T] &\simeq  \frac{1}{2}a_T(-q) \left[ \frac{e_*^2}{8\pi} \vert q \vert + \sigma_{xx}^{\rm cf}|\omega|  + \mathcal O(\lambda^2 q^2 )  \right] a_T(q)
	\label{leff}.
	\end{align}
	We have neglected the constant $\mathcal O(B^2)$ correction to $\mathcal L'_{gauge}$ after integrating out $a_\tau$.\footnote{Note that the term $\mathcal{L}[A]=-\frac{i}{4\pi}\frac{1}{2m} AdA$ is present in addition and determines the dictionary between the conductivity of CFs and the EM charge.}  Thus, since the CS term is subleading in the presence of Coulomb interactions at criticality, it immediately follows that fractional and integer QH transitions are in the same universality class, with identical critical exponents $\nu$, CF conductivity tensor, and dynamical scaling set by an overdamped transverse gauge boson.
	Again, the robustness of this conclusion rests on the fact that the critical point has a metallic description in terms of delocalized composite fermion states.  The notion of superuniversality in previous studies had remained conjectural\cite{Kivelson1992} (for a more recent discussion, see \onlinecite{Hui2019}); here we have provided direct arguments based on the presence of $1/r$ interactions.
	
	The CS term of the emergent gauge field is irrelevant for the properties of the CFs, however, it determines the dictionaries between the CF and electromagnetic conductivities. The resulting electron conductivities are consistent with the correspondence rules conjectured in Ref. \onlinecite{Kivelson1992}
	
	The arguments above are easily repeated in the case of short-range interactions, with 
	\begin{equation}
	\mathcal L'_{\rm int} = \frac{U_0}{16\pi}  q^2   a_T(-q)   a_T (q)
	\end{equation}
	In this instance, gauge fluctuation effects are stronger and the interplay between disorder and interactions remains more subtle.  Integrating out $a_\tau$ and invoking the Kubo formula, 
	\begin{align}
	\mathcal L'_{\rm eff}[a_T] &= \frac{1}{2}  a_T( -q)\left[ \left(\frac{U_0}{8 \pi}  +  \frac{\lambda^2}{N_F} \right) q^2 + |\omega| \sigma_{xx}^{\rm cf} \right] a_T( q)
	\end{align} 
	Thus, we can expect that at the critical point, $z=2$.  However, the CS term does not decouple in this case, and more sophisticated analysis is needed to conclude whether or not the transitions are superuniversal.

	\section{Summary\label{sec:summary}}	
	We have studied QH to insulator transitions in the composite fermion representation, which provides a new perspective on the problem.  In the CF representation, delocalized states occur at all energies at criticality.  We have concluded that $z=1(z=2)$ for $1/r$(short-range) interactions.  We have also shown that with $1/r$ interactions, the transitions are superuniversal.  Moreover, in the Dirac CF theory, the Hall conductance is not renormalized at criticality by gauge fluctuations.  Further study including non-perturbative effects of disorder are needed to determine the localization length exponent $\nu$.  
	We will report on these developments elsewhere.  On the experimental front, further studies of QH transitions with gates will be of fundamental importance to shed light on dynamical scaling laws and superuniversality in the presence of screened Coulomb interactions.  
	We have provided an explicit proof that the transitions with $1/r$ interactions are superuniversal. Our conclusions go beyond previous conjectures based solely on the mean-field aproximation and take into account gauge fluctuations. With screened interactions, arguments for superuniversality remain more subtle, and will likely produce distict finite temperature crossovers.  We shall report on such phenomena in future studies.
	
	\acknowledgements
	
	We thank I. Burmistrov, E. Fradkin, A. Kamenev, Y.-B. Kim, S. Kivelson, D.-H. Lee, and M. Mulligan for fruitful discussions.  This work was supported in part by the US Department of Energy, Office of Basic Energy Sciences, Division of Materials Sciences and Engineering, under contract number DE-AC02-76SF00515. P.K. was supported in part by DOE BES Grant
	No. DE-SC0002140.
	
	\bibliography{bigbib}

\begin{thebibliography}{43}%
\makeatletter
\providecommand \@ifxundefined [1]{%
 \@ifx{#1\undefined}
}%
\providecommand \@ifnum [1]{%
 \ifnum #1\expandafter \@firstoftwo
 \else \expandafter \@secondoftwo
 \fi
}%
\providecommand \@ifx [1]{%
 \ifx #1\expandafter \@firstoftwo
 \else \expandafter \@secondoftwo
 \fi
}%
\providecommand \natexlab [1]{#1}%
\providecommand \enquote  [1]{``#1''}%
\providecommand \bibnamefont  [1]{#1}%
\providecommand \bibfnamefont [1]{#1}%
\providecommand \citenamefont [1]{#1}%
\providecommand \href@noop [0]{\@secondoftwo}%
\providecommand \href [0]{\begingroup \@sanitize@url \@href}%
\providecommand \@href[1]{\@@startlink{#1}\@@href}%
\providecommand \@@href[1]{\endgroup#1\@@endlink}%
\providecommand \@sanitize@url [0]{\catcode `\\12\catcode `\$12\catcode
  `\&12\catcode `\#12\catcode `\^12\catcode `\_12\catcode `\%12\relax}%
\providecommand \@@startlink[1]{}%
\providecommand \@@endlink[0]{}%
\providecommand \url  [0]{\begingroup\@sanitize@url \@url }%
\providecommand \@url [1]{\endgroup\@href {#1}{\urlprefix }}%
\providecommand \urlprefix  [0]{URL }%
\providecommand \Eprint [0]{\href }%
\providecommand \doibase [0]{http://dx.doi.org/}%
\providecommand \selectlanguage [0]{\@gobble}%
\providecommand \bibinfo  [0]{\@secondoftwo}%
\providecommand \bibfield  [0]{\@secondoftwo}%
\providecommand \translation [1]{[#1]}%
\providecommand \BibitemOpen [0]{}%
\providecommand \bibitemStop [0]{}%
\providecommand \bibitemNoStop [0]{.\EOS\space}%
\providecommand \EOS [0]{\spacefactor3000\relax}%
\providecommand \BibitemShut  [1]{\csname bibitem#1\endcsname}%
\let\auto@bib@innerbib\@empty
\bibitem [{\citenamefont {Sondhi}\ \emph {et~al.}(1997)\citenamefont {Sondhi},
  \citenamefont {Girvin}, \citenamefont {Carini},\ and\ \citenamefont
  {Shahar}}]{SondhiGirvinCariniShahar}%
  \BibitemOpen
  \bibfield  {author} {\bibinfo {author} {\bibfnamefont {S.~L.}\ \bibnamefont
  {Sondhi}}, \bibinfo {author} {\bibfnamefont {S.~M.}\ \bibnamefont {Girvin}},
  \bibinfo {author} {\bibfnamefont {J.~P.}\ \bibnamefont {Carini}}, \ and\
  \bibinfo {author} {\bibfnamefont {D.}~\bibnamefont {Shahar}},\ }\href
  {\doibase 10.1103/RevModPhys.69.315} {\bibfield  {journal} {\bibinfo
  {journal} {Rev. Mod. Phys.}\ }\textbf {\bibinfo {volume} {69}},\ \bibinfo
  {pages} {315} (\bibinfo {year} {1997})}\BibitemShut {NoStop}%
\bibitem [{\citenamefont {Wang}\ \emph {et~al.}(2000)\citenamefont {Wang},
  \citenamefont {Fisher}, \citenamefont {Girvin},\ and\ \citenamefont
  {Chalker}}]{Wang2000}%
  \BibitemOpen
  \bibfield  {author} {\bibinfo {author} {\bibfnamefont {Z.}~\bibnamefont
  {Wang}}, \bibinfo {author} {\bibfnamefont {M.~P.~A.}\ \bibnamefont {Fisher}},
  \bibinfo {author} {\bibfnamefont {S.~M.}\ \bibnamefont {Girvin}}, \ and\
  \bibinfo {author} {\bibfnamefont {J.~T.}\ \bibnamefont {Chalker}},\ }\href
  {\doibase 10.1103/PhysRevB.61.8326} {\bibfield  {journal} {\bibinfo
  {journal} {Phys. Rev. B}\ }\textbf {\bibinfo {volume} {61}},\ \bibinfo
  {pages} {8326} (\bibinfo {year} {2000})}\BibitemShut {NoStop}%
\bibitem [{\citenamefont {Engel}\ \emph {et~al.}(1993)\citenamefont {Engel},
  \citenamefont {Shahar}, \citenamefont {Kurdak},\ and\ \citenamefont
  {Tsui}}]{Engel1993}%
  \BibitemOpen
  \bibfield  {author} {\bibinfo {author} {\bibfnamefont {L.~W.}\ \bibnamefont
  {Engel}}, \bibinfo {author} {\bibfnamefont {D.}~\bibnamefont {Shahar}},
  \bibinfo {author} {\bibfnamefont {i.~m.~c.}\ \bibnamefont {Kurdak}}, \ and\
  \bibinfo {author} {\bibfnamefont {D.~C.}\ \bibnamefont {Tsui}},\ }\href
  {\doibase 10.1103/PhysRevLett.71.2638} {\bibfield  {journal} {\bibinfo
  {journal} {Phys. Rev. Lett.}\ }\textbf {\bibinfo {volume} {71}},\ \bibinfo
  {pages} {2638} (\bibinfo {year} {1993})}\BibitemShut {NoStop}%
\bibitem [{\citenamefont {Wei}\ \emph {et~al.}(1994)\citenamefont {Wei},
  \citenamefont {Engel},\ and\ \citenamefont {Tsui}}]{Wei1994}%
  \BibitemOpen
  \bibfield  {author} {\bibinfo {author} {\bibfnamefont {H.~P.}\ \bibnamefont
  {Wei}}, \bibinfo {author} {\bibfnamefont {L.~W.}\ \bibnamefont {Engel}}, \
  and\ \bibinfo {author} {\bibfnamefont {D.~C.}\ \bibnamefont {Tsui}},\ }\href
  {\doibase 10.1103/PhysRevB.50.14609} {\bibfield  {journal} {\bibinfo
  {journal} {Phys. Rev. B}\ }\textbf {\bibinfo {volume} {50}},\ \bibinfo
  {pages} {14609} (\bibinfo {year} {1994})}\BibitemShut {NoStop}%
\bibitem [{\citenamefont {Li}\ \emph {et~al.}(2009)\citenamefont {Li},
  \citenamefont {Vicente}, \citenamefont {Xia}, \citenamefont {Pan},
  \citenamefont {Tsui}, \citenamefont {Pfeiffer},\ and\ \citenamefont
  {West}}]{Li2009}%
  \BibitemOpen
  \bibfield  {author} {\bibinfo {author} {\bibfnamefont {W.}~\bibnamefont
  {Li}}, \bibinfo {author} {\bibfnamefont {C.~L.}\ \bibnamefont {Vicente}},
  \bibinfo {author} {\bibfnamefont {J.~S.}\ \bibnamefont {Xia}}, \bibinfo
  {author} {\bibfnamefont {W.}~\bibnamefont {Pan}}, \bibinfo {author}
  {\bibfnamefont {D.~C.}\ \bibnamefont {Tsui}}, \bibinfo {author}
  {\bibfnamefont {L.~N.}\ \bibnamefont {Pfeiffer}}, \ and\ \bibinfo {author}
  {\bibfnamefont {K.~W.}\ \bibnamefont {West}},\ }\href {\doibase
  10.1103/PhysRevLett.102.216801} {\bibfield  {journal} {\bibinfo  {journal}
  {Phys. Rev. Lett.}\ }\textbf {\bibinfo {volume} {102}},\ \bibinfo {pages}
  {216801} (\bibinfo {year} {2009})}\BibitemShut {NoStop}%
\bibitem [{\citenamefont {Halperin}\ \emph {et~al.}(1993)\citenamefont
  {Halperin}, \citenamefont {Lee},\ and\ \citenamefont {Read}}]{Halperin1993}%
  \BibitemOpen
  \bibfield  {author} {\bibinfo {author} {\bibfnamefont {B.~I.}\ \bibnamefont
  {Halperin}}, \bibinfo {author} {\bibfnamefont {P.~A.}\ \bibnamefont {Lee}}, \
  and\ \bibinfo {author} {\bibfnamefont {N.}~\bibnamefont {Read}},\ }\href
  {\doibase 10.1103/PhysRevB.47.7312} {\bibfield  {journal} {\bibinfo
  {journal} {Phys. Rev. B}\ }\textbf {\bibinfo {volume} {47}},\ \bibinfo
  {pages} {7312} (\bibinfo {year} {1993})}\BibitemShut {NoStop}%
\bibitem [{\citenamefont {Son}(2015)}]{Son2015}%
  \BibitemOpen
  \bibfield  {author} {\bibinfo {author} {\bibfnamefont {D.~T.}\ \bibnamefont
  {Son}},\ }\href {\doibase 10.1103/PhysRevX.5.031027} {\bibfield  {journal}
  {\bibinfo  {journal} {Phys. Rev. X}\ }\textbf {\bibinfo {volume} {5}},\
  \bibinfo {pages} {031027} (\bibinfo {year} {2015})}\BibitemShut {NoStop}%
\bibitem [{\citenamefont {Kumar}\ \emph
  {et~al.}(2019{\natexlab{a}})\citenamefont {Kumar}, \citenamefont {Kim},\ and\
  \citenamefont {Raghu}}]{Kumar2019b}%
  \BibitemOpen
  \bibfield  {author} {\bibinfo {author} {\bibfnamefont {P.}~\bibnamefont
  {Kumar}}, \bibinfo {author} {\bibfnamefont {Y.~B.}\ \bibnamefont {Kim}}, \
  and\ \bibinfo {author} {\bibfnamefont {S.}~\bibnamefont {Raghu}},\ }\href
  {\doibase 10.1103/PhysRevB.100.235124} {\bibfield  {journal} {\bibinfo
  {journal} {Phys. Rev. B}\ }\textbf {\bibinfo {volume} {100}},\ \bibinfo
  {pages} {235124} (\bibinfo {year} {2019}{\natexlab{a}})}\BibitemShut
  {NoStop}%
\bibitem [{\citenamefont {Huang}\ \emph {et~al.}(2021)\citenamefont {Huang},
  \citenamefont {Raghu},\ and\ \citenamefont {Kumar}}]{Huang2021}%
  \BibitemOpen
  \bibfield  {author} {\bibinfo {author} {\bibfnamefont {K.~S.}\ \bibnamefont
  {Huang}}, \bibinfo {author} {\bibfnamefont {S.}~\bibnamefont {Raghu}}, \ and\
  \bibinfo {author} {\bibfnamefont {P.}~\bibnamefont {Kumar}},\ }\href
  {\doibase 10.1103/PhysRevLett.126.056802} {\bibfield  {journal} {\bibinfo
  {journal} {Phys. Rev. Lett.}\ }\textbf {\bibinfo {volume} {126}},\ \bibinfo
  {pages} {056802} (\bibinfo {year} {2021})}\BibitemShut {NoStop}%
\bibitem [{\citenamefont {Engel}\ \emph {et~al.}(1990)\citenamefont {Engel},
  \citenamefont {Wei}, \citenamefont {Tsui},\ and\ \citenamefont
  {Shayegan}}]{Engel1990}%
  \BibitemOpen
  \bibfield  {author} {\bibinfo {author} {\bibfnamefont {L.}~\bibnamefont
  {Engel}}, \bibinfo {author} {\bibfnamefont {H.}~\bibnamefont {Wei}}, \bibinfo
  {author} {\bibfnamefont {D.}~\bibnamefont {Tsui}}, \ and\ \bibinfo {author}
  {\bibfnamefont {M.}~\bibnamefont {Shayegan}},\ }\href {\doibase
  https://doi.org/10.1016/0039-6028(90)90820-X} {\bibfield  {journal} {\bibinfo
   {journal} {Surface Science}\ }\textbf {\bibinfo {volume} {229}},\ \bibinfo
  {pages} {13 } (\bibinfo {year} {1990})}\BibitemShut {NoStop}%
\bibitem [{\citenamefont {Kivelson}\ \emph {et~al.}(1992)\citenamefont
  {Kivelson}, \citenamefont {Lee},\ and\ \citenamefont {Zhang}}]{Kivelson1992}%
  \BibitemOpen
  \bibfield  {author} {\bibinfo {author} {\bibfnamefont {S.}~\bibnamefont
  {Kivelson}}, \bibinfo {author} {\bibfnamefont {D.-H.}\ \bibnamefont {Lee}}, \
  and\ \bibinfo {author} {\bibfnamefont {S.-C.}\ \bibnamefont {Zhang}},\ }\href
  {\doibase 10.1103/PhysRevB.46.2223} {\bibfield  {journal} {\bibinfo
  {journal} {Phys. Rev. B}\ }\textbf {\bibinfo {volume} {46}},\ \bibinfo
  {pages} {2223} (\bibinfo {year} {1992})}\BibitemShut {NoStop}%
\bibitem [{\citenamefont {Huo}\ \emph {et~al.}(1993)\citenamefont {Huo},
  \citenamefont {Hetzel},\ and\ \citenamefont {Bhatt}}]{PhysRevLett.70.481}%
  \BibitemOpen
  \bibfield  {author} {\bibinfo {author} {\bibfnamefont {Y.}~\bibnamefont
  {Huo}}, \bibinfo {author} {\bibfnamefont {R.~E.}\ \bibnamefont {Hetzel}}, \
  and\ \bibinfo {author} {\bibfnamefont {R.~N.}\ \bibnamefont {Bhatt}},\ }\href
  {\doibase 10.1103/PhysRevLett.70.481} {\bibfield  {journal} {\bibinfo
  {journal} {Phys. Rev. Lett.}\ }\textbf {\bibinfo {volume} {70}},\ \bibinfo
  {pages} {481} (\bibinfo {year} {1993})}\BibitemShut {NoStop}%
\bibitem [{\citenamefont {Shahar}\ \emph {et~al.}(1995)\citenamefont {Shahar},
  \citenamefont {Tsui}, \citenamefont {Shayegan}, \citenamefont {Bhatt},\ and\
  \citenamefont {Cunningham}}]{Shahar1995}%
  \BibitemOpen
  \bibfield  {author} {\bibinfo {author} {\bibfnamefont {D.}~\bibnamefont
  {Shahar}}, \bibinfo {author} {\bibfnamefont {D.~C.}\ \bibnamefont {Tsui}},
  \bibinfo {author} {\bibfnamefont {M.}~\bibnamefont {Shayegan}}, \bibinfo
  {author} {\bibfnamefont {R.~N.}\ \bibnamefont {Bhatt}}, \ and\ \bibinfo
  {author} {\bibfnamefont {J.~E.}\ \bibnamefont {Cunningham}},\ }\href
  {\doibase 10.1103/PhysRevLett.74.4511} {\bibfield  {journal} {\bibinfo
  {journal} {Phys. Rev. Lett.}\ }\textbf {\bibinfo {volume} {74}},\ \bibinfo
  {pages} {4511} (\bibinfo {year} {1995})}\BibitemShut {NoStop}%
\bibitem [{\citenamefont {{Seiberg}}\ \emph {et~al.}(2016)\citenamefont
  {{Seiberg}}, \citenamefont {{Senthil}}, \citenamefont {{Wang}},\ and\
  \citenamefont {{Witten}}}]{Seiberg:2016gmd}%
  \BibitemOpen
  \bibfield  {author} {\bibinfo {author} {\bibfnamefont {N.}~\bibnamefont
  {{Seiberg}}}, \bibinfo {author} {\bibfnamefont {T.}~\bibnamefont
  {{Senthil}}}, \bibinfo {author} {\bibfnamefont {C.}~\bibnamefont {{Wang}}}, \
  and\ \bibinfo {author} {\bibfnamefont {E.}~\bibnamefont {{Witten}}},\ }\href
  {\doibase 10.1016/j.aop.2016.08.007} {\bibfield  {journal} {\bibinfo
  {journal} {Annals of Physics}\ }\textbf {\bibinfo {volume} {374}},\ \bibinfo
  {pages} {395} (\bibinfo {year} {2016})},\ \Eprint
  {http://arxiv.org/abs/1606.01989} {arXiv:1606.01989 [hep-th]} \BibitemShut
  {NoStop}%
\bibitem [{\citenamefont {Redlich}(1984)}]{Redlichparitylong}%
  \BibitemOpen
  \bibfield  {author} {\bibinfo {author} {\bibfnamefont {A.~N.}\ \bibnamefont
  {Redlich}},\ }\href {\doibase 10.1103/PhysRevD.29.2366} {\bibfield  {journal}
  {\bibinfo  {journal} {Phys. Rev.}\ }\textbf {\bibinfo {volume} {D29}},\
  \bibinfo {pages} {2366} (\bibinfo {year} {1984})}\BibitemShut {NoStop}%
\bibitem [{\citenamefont {Son}\ \emph {et~al.}(2019)\citenamefont {Son},
  \citenamefont {Chen},\ and\ \citenamefont {Raghu}}]{Son2019}%
  \BibitemOpen
  \bibfield  {author} {\bibinfo {author} {\bibfnamefont {J.~H.}\ \bibnamefont
  {Son}}, \bibinfo {author} {\bibfnamefont {J.-Y.}\ \bibnamefont {Chen}}, \
  and\ \bibinfo {author} {\bibfnamefont {S.}~\bibnamefont {Raghu}},\ }\href
  {\doibase 10.1007/jhep06(2019)038} {\bibfield  {journal} {\bibinfo  {journal}
  {Journal of High Energy Physics}\ }\textbf {\bibinfo {volume} {2019}}
  (\bibinfo {year} {2019}),\ 10.1007/jhep06(2019)038}\BibitemShut {NoStop}%
\bibitem [{\citenamefont {Laughlin}(1981)}]{Laughlin1981}%
  \BibitemOpen
  \bibfield  {author} {\bibinfo {author} {\bibfnamefont {R.~B.}\ \bibnamefont
  {Laughlin}},\ }\href {\doibase 10.1103/PhysRevB.23.5632} {\bibfield
  {journal} {\bibinfo  {journal} {Phys. Rev. B}\ }\textbf {\bibinfo {volume}
  {23}},\ \bibinfo {pages} {5632} (\bibinfo {year} {1981})}\BibitemShut
  {NoStop}%
\bibitem [{\citenamefont {Aharonov}\ and\ \citenamefont
  {Casher}(1979)}]{Aharonov1979}%
  \BibitemOpen
  \bibfield  {author} {\bibinfo {author} {\bibfnamefont {Y.}~\bibnamefont
  {Aharonov}}\ and\ \bibinfo {author} {\bibfnamefont {A.}~\bibnamefont
  {Casher}},\ }\href {\doibase 10.1103/PhysRevA.19.2461} {\bibfield  {journal}
  {\bibinfo  {journal} {Phys. Rev. A}\ }\textbf {\bibinfo {volume} {19}},\
  \bibinfo {pages} {2461} (\bibinfo {year} {1979})}\BibitemShut {NoStop}%
\bibitem [{\citenamefont {{Goldman}}\ \emph {et~al.}(2017)\citenamefont
  {{Goldman}}, \citenamefont {{Mulligan}}, \citenamefont {{Raghu}},
  \citenamefont {{Torroba}},\ and\ \citenamefont
  {{Zimet}}}]{2017PhRvB..96x5140G}%
  \BibitemOpen
  \bibfield  {author} {\bibinfo {author} {\bibfnamefont {H.}~\bibnamefont
  {{Goldman}}}, \bibinfo {author} {\bibfnamefont {M.}~\bibnamefont
  {{Mulligan}}}, \bibinfo {author} {\bibfnamefont {S.}~\bibnamefont {{Raghu}}},
  \bibinfo {author} {\bibfnamefont {G.}~\bibnamefont {{Torroba}}}, \ and\
  \bibinfo {author} {\bibfnamefont {M.}~\bibnamefont {{Zimet}}},\ }\href
  {\doibase 10.1103/PhysRevB.96.245140} {\bibfield  {journal} {\bibinfo
  {journal} {Phys. Rev. B}\ }\textbf {\bibinfo {volume} {96}},\ \bibinfo {eid}
  {245140} (\bibinfo {year} {2017})},\ \Eprint
  {http://arxiv.org/abs/1709.07005} {arXiv:1709.07005 [cond-mat.str-el]}
  \BibitemShut {NoStop}%
\bibitem [{\citenamefont {Kumar}\ \emph {et~al.}(2018)\citenamefont {Kumar},
  \citenamefont {Mulligan},\ and\ \citenamefont {Raghu}}]{Kumarsusy}%
  \BibitemOpen
  \bibfield  {author} {\bibinfo {author} {\bibfnamefont {P.}~\bibnamefont
  {Kumar}}, \bibinfo {author} {\bibfnamefont {M.}~\bibnamefont {Mulligan}}, \
  and\ \bibinfo {author} {\bibfnamefont {S.}~\bibnamefont {Raghu}},\ }\href
  {\doibase 10.1103/PhysRevB.98.115105} {\bibfield  {journal} {\bibinfo
  {journal} {Phys. Rev. B}\ }\textbf {\bibinfo {volume} {98}},\ \bibinfo
  {pages} {115105} (\bibinfo {year} {2018})}\BibitemShut {NoStop}%
\bibitem [{\citenamefont {Liu}\ and\ \citenamefont {Bhatt}(2016)}]{Liu2016}%
  \BibitemOpen
  \bibfield  {author} {\bibinfo {author} {\bibfnamefont {Z.}~\bibnamefont
  {Liu}}\ and\ \bibinfo {author} {\bibfnamefont {R.~N.}\ \bibnamefont
  {Bhatt}},\ }\href {\doibase 10.1103/PhysRevLett.117.206801} {\bibfield
  {journal} {\bibinfo  {journal} {Phys. Rev. Lett.}\ }\textbf {\bibinfo
  {volume} {117}},\ \bibinfo {pages} {206801} (\bibinfo {year}
  {2016})}\BibitemShut {NoStop}%
\bibitem [{\citenamefont {Goldman}\ and\ \citenamefont
  {Fradkin}(2018)}]{Goldman2018}%
  \BibitemOpen
  \bibfield  {author} {\bibinfo {author} {\bibfnamefont {H.}~\bibnamefont
  {Goldman}}\ and\ \bibinfo {author} {\bibfnamefont {E.}~\bibnamefont
  {Fradkin}},\ }\href {\doibase 10.1103/PhysRevB.98.165137} {\bibfield
  {journal} {\bibinfo  {journal} {Phys. Rev. B}\ }\textbf {\bibinfo {volume}
  {98}},\ \bibinfo {pages} {165137} (\bibinfo {year} {2018})}\BibitemShut
  {NoStop}%
\bibitem [{\citenamefont {Belitz}\ and\ \citenamefont
  {Kirkpatrick}(1994)}]{Belitz1994}%
  \BibitemOpen
  \bibfield  {author} {\bibinfo {author} {\bibfnamefont {D.}~\bibnamefont
  {Belitz}}\ and\ \bibinfo {author} {\bibfnamefont {T.~R.}\ \bibnamefont
  {Kirkpatrick}},\ }\href {\doibase 10.1103/RevModPhys.66.261} {\bibfield
  {journal} {\bibinfo  {journal} {Rev. Mod. Phys.}\ }\textbf {\bibinfo {volume}
  {66}},\ \bibinfo {pages} {261} (\bibinfo {year} {1994})}\BibitemShut
  {NoStop}%
\bibitem [{\citenamefont {Nayak}()}]{Nayaklectures}%
  \BibitemOpen
  \bibfield  {author} {\bibinfo {author} {\bibfnamefont {C.}~\bibnamefont
  {Nayak}},\ }\href@noop {} {\enquote {\bibinfo {title} {{Quantum Condensed
  Matter Physics - Lecture Notes}},}\ }\bibinfo {note}
  {Unpublished}\BibitemShut {NoStop}%
\bibitem [{\citenamefont {Kamenev}(2011)}]{kamenev2011field}%
  \BibitemOpen
  \bibfield  {author} {\bibinfo {author} {\bibfnamefont {A.}~\bibnamefont
  {Kamenev}},\ }\href@noop {} {\emph {\bibinfo {title} {Field theory of
  non-equilibrium systems}}}\ (\bibinfo  {publisher} {Cambridge University
  Press},\ \bibinfo {address} {Cambridge New York},\ \bibinfo {year}
  {2011})\BibitemShut {NoStop}%
\bibitem [{\citenamefont {Pruisken}\ \emph {et~al.}(1999)\citenamefont
  {Pruisken}, \citenamefont {Baranov},\ and\ \citenamefont {\ifmmode
  \check{S}\else \v{S}\fi{}kori\ifmmode~\acute{c}\else
  \'{c}\fi{}}}]{Pruisken1999}%
  \BibitemOpen
  \bibfield  {author} {\bibinfo {author} {\bibfnamefont {A.~M.~M.}\
  \bibnamefont {Pruisken}}, \bibinfo {author} {\bibfnamefont {M.~A.}\
  \bibnamefont {Baranov}}, \ and\ \bibinfo {author} {\bibfnamefont
  {B.}~\bibnamefont {\ifmmode \check{S}\else
  \v{S}\fi{}kori\ifmmode~\acute{c}\else \'{c}\fi{}}},\ }\href {\doibase
  10.1103/PhysRevB.60.16807} {\bibfield  {journal} {\bibinfo  {journal} {Phys.
  Rev. B}\ }\textbf {\bibinfo {volume} {60}},\ \bibinfo {pages} {16807}
  (\bibinfo {year} {1999})}\BibitemShut {NoStop}%
\bibitem [{\citenamefont {Pruisken}(1984)}]{Pruisken1984}%
  \BibitemOpen
  \bibfield  {author} {\bibinfo {author} {\bibfnamefont {A.}~\bibnamefont
  {Pruisken}},\ }\href {\doibase 10.1016/0550-3213(84)90101-9} {\bibfield
  {journal} {\bibinfo  {journal} {Nuclear Physics B}\ }\textbf {\bibinfo
  {volume} {235}},\ \bibinfo {pages} {277 } (\bibinfo {year}
  {1984})}\BibitemShut {NoStop}%
\bibitem [{\citenamefont {K\"onig}\ \emph {et~al.}(2014)\citenamefont
  {K\"onig}, \citenamefont {Ostrovsky}, \citenamefont {Protopopov},
  \citenamefont {Gornyi}, \citenamefont {Burmistrov},\ and\ \citenamefont
  {Mirlin}}]{Mirlin2014}%
  \BibitemOpen
  \bibfield  {author} {\bibinfo {author} {\bibfnamefont {E.~J.}\ \bibnamefont
  {K\"onig}}, \bibinfo {author} {\bibfnamefont {P.~M.}\ \bibnamefont
  {Ostrovsky}}, \bibinfo {author} {\bibfnamefont {I.~V.}\ \bibnamefont
  {Protopopov}}, \bibinfo {author} {\bibfnamefont {I.~V.}\ \bibnamefont
  {Gornyi}}, \bibinfo {author} {\bibfnamefont {I.~S.}\ \bibnamefont
  {Burmistrov}}, \ and\ \bibinfo {author} {\bibfnamefont {A.~D.}\ \bibnamefont
  {Mirlin}},\ }\href {\doibase 10.1103/PhysRevB.90.165435} {\bibfield
  {journal} {\bibinfo  {journal} {Phys. Rev. B}\ }\textbf {\bibinfo {volume}
  {90}},\ \bibinfo {pages} {165435} (\bibinfo {year} {2014})}\BibitemShut
  {NoStop}%
\bibitem [{Note1()}]{Note1}%
  \BibitemOpen
  \bibinfo {note} {By time-reversal symmetry, we mean that the contribution
  from the light Dirac fermions to the Hall conductance is identically zero.
  For further details, see Ref. \protect \rev@citealpnum {Son2015}}\BibitemShut
  {NoStop}%
\bibitem [{\citenamefont {Mitra}\ and\ \citenamefont
  {Mulligan}(2019)}]{Mitra2019}%
  \BibitemOpen
  \bibfield  {author} {\bibinfo {author} {\bibfnamefont {A.}~\bibnamefont
  {Mitra}}\ and\ \bibinfo {author} {\bibfnamefont {M.}~\bibnamefont
  {Mulligan}},\ }\href {\doibase 10.1103/PhysRevB.100.165122} {\bibfield
  {journal} {\bibinfo  {journal} {Phys. Rev. B}\ }\textbf {\bibinfo {volume}
  {100}},\ \bibinfo {pages} {165122} (\bibinfo {year} {2019})}\BibitemShut
  {NoStop}%
\bibitem [{Note2()}]{Note2}%
  \BibitemOpen
  \bibinfo {note} {In the expression for $B_j$ in Eq. \ref {Bj}, derivatives
  are not operators; they only act directly on $u$.}\BibitemShut {Stop}%
\bibitem [{\citenamefont {Fujikawa}(1979)}]{Fujikawa1979}%
  \BibitemOpen
  \bibfield  {author} {\bibinfo {author} {\bibfnamefont {K.}~\bibnamefont
  {Fujikawa}},\ }\href {\doibase 10.1103/PhysRevLett.42.1195} {\bibfield
  {journal} {\bibinfo  {journal} {Phys. Rev. Lett.}\ }\textbf {\bibinfo
  {volume} {42}},\ \bibinfo {pages} {1195} (\bibinfo {year}
  {1979})}\BibitemShut {NoStop}%
\bibitem [{\citenamefont {Chow}\ \emph {et~al.}(1996)\citenamefont {Chow},
  \citenamefont {Wei}, \citenamefont {Girvin},\ and\ \citenamefont
  {Shayegan}}]{Chow1996}%
  \BibitemOpen
  \bibfield  {author} {\bibinfo {author} {\bibfnamefont {E.}~\bibnamefont
  {Chow}}, \bibinfo {author} {\bibfnamefont {H.~P.}\ \bibnamefont {Wei}},
  \bibinfo {author} {\bibfnamefont {S.~M.}\ \bibnamefont {Girvin}}, \ and\
  \bibinfo {author} {\bibfnamefont {M.}~\bibnamefont {Shayegan}},\ }\href
  {\doibase 10.1103/PhysRevLett.77.1143} {\bibfield  {journal} {\bibinfo
  {journal} {Phys. Rev. Lett.}\ }\textbf {\bibinfo {volume} {77}},\ \bibinfo
  {pages} {1143} (\bibinfo {year} {1996})}\BibitemShut {NoStop}%
\bibitem [{Note3()}]{Note3}%
  \BibitemOpen
  \bibinfo {note} {Such plasmon type reasoning was also proposed to be the
  source of $z=1$ scaling for superconductor to insulator transition in Ref.
  \protect \rev@citealpnum {Fisher1990}.}\BibitemShut {Stop}%
\bibitem [{\citenamefont {Pruisken}\ and\ \citenamefont
  {Burmistrov}(2009)}]{Pruisken2009comment}%
  \BibitemOpen
  \bibfield  {author} {\bibinfo {author} {\bibfnamefont {A.~M.~M.}\
  \bibnamefont {Pruisken}}\ and\ \bibinfo {author} {\bibfnamefont {I.~S.}\
  \bibnamefont {Burmistrov}},\ }\href {https://arxiv.org/abs/0907.0356} {\
  (\bibinfo {year} {2009})},\ \Eprint {http://arxiv.org/abs/0907.0356}
  {arXiv:0907.0356 [cond-mat.mes-hall]} \BibitemShut {NoStop}%
\bibitem [{Note4()}]{Note4}%
  \BibitemOpen
  \bibinfo {note} {Note that the term $\protect \mathcal {L}[A]=-\protect \frac
  {i}{4\pi }\protect \frac {1}{2m} AdA$ is present in addition and determines
  the dictionary between the conductivity of CFs and the EM
  charge.}\BibitemShut {Stop}%
\bibitem [{\citenamefont {Hui}\ \emph {et~al.}(2019)\citenamefont {Hui},
  \citenamefont {Kim},\ and\ \citenamefont {Mulligan}}]{Hui2019}%
  \BibitemOpen
  \bibfield  {author} {\bibinfo {author} {\bibfnamefont {A.}~\bibnamefont
  {Hui}}, \bibinfo {author} {\bibfnamefont {E.-A.}\ \bibnamefont {Kim}}, \ and\
  \bibinfo {author} {\bibfnamefont {M.}~\bibnamefont {Mulligan}},\ }\href
  {\doibase 10.1103/PhysRevB.99.125135} {\bibfield  {journal} {\bibinfo
  {journal} {Phys. Rev. B}\ }\textbf {\bibinfo {volume} {99}},\ \bibinfo
  {pages} {125135} (\bibinfo {year} {2019})}\BibitemShut {NoStop}%
\bibitem [{\citenamefont {Fisher}\ \emph {et~al.}(1990)\citenamefont {Fisher},
  \citenamefont {Grinstein},\ and\ \citenamefont {Girvin}}]{Fisher1990}%
  \BibitemOpen
  \bibfield  {author} {\bibinfo {author} {\bibfnamefont {M.~P.~A.}\
  \bibnamefont {Fisher}}, \bibinfo {author} {\bibfnamefont {G.}~\bibnamefont
  {Grinstein}}, \ and\ \bibinfo {author} {\bibfnamefont {S.~M.}\ \bibnamefont
  {Girvin}},\ }\href {\doibase 10.1103/PhysRevLett.64.587} {\bibfield
  {journal} {\bibinfo  {journal} {Phys. Rev. Lett.}\ }\textbf {\bibinfo
  {volume} {64}},\ \bibinfo {pages} {587} (\bibinfo {year} {1990})}\BibitemShut
  {NoStop}%
\bibitem [{\citenamefont {Wang}\ \emph {et~al.}(2017)\citenamefont {Wang},
  \citenamefont {Cooper}, \citenamefont {Halperin},\ and\ \citenamefont
  {Stern}}]{Wang2017}%
  \BibitemOpen
  \bibfield  {author} {\bibinfo {author} {\bibfnamefont {C.}~\bibnamefont
  {Wang}}, \bibinfo {author} {\bibfnamefont {N.~R.}\ \bibnamefont {Cooper}},
  \bibinfo {author} {\bibfnamefont {B.~I.}\ \bibnamefont {Halperin}}, \ and\
  \bibinfo {author} {\bibfnamefont {A.}~\bibnamefont {Stern}},\ }\href
  {\doibase 10.1103/PhysRevX.7.031029} {\bibfield  {journal} {\bibinfo
  {journal} {Phys. Rev. X}\ }\textbf {\bibinfo {volume} {7}},\ \bibinfo {pages}
  {031029} (\bibinfo {year} {2017})}\BibitemShut {NoStop}%
\bibitem [{\citenamefont {Kumar}\ \emph
  {et~al.}(2019{\natexlab{b}})\citenamefont {Kumar}, \citenamefont {Raghu},\
  and\ \citenamefont {Mulligan}}]{KumarTrivialPaper}%
  \BibitemOpen
  \bibfield  {author} {\bibinfo {author} {\bibfnamefont {P.}~\bibnamefont
  {Kumar}}, \bibinfo {author} {\bibfnamefont {S.}~\bibnamefont {Raghu}}, \ and\
  \bibinfo {author} {\bibfnamefont {M.}~\bibnamefont {Mulligan}},\ }\href
  {\doibase 10.1103/PhysRevB.99.235114} {\bibfield  {journal} {\bibinfo
  {journal} {Phys. Rev. B}\ }\textbf {\bibinfo {volume} {99}},\ \bibinfo
  {pages} {235114} (\bibinfo {year} {2019}{\natexlab{b}})}\BibitemShut
  {NoStop}%
\bibitem [{\citenamefont {Ludwig}\ \emph {et~al.}(2008)\citenamefont {Ludwig},
  \citenamefont {Gornyi}, \citenamefont {Mirlin},\ and\ \citenamefont
  {W\"olfle}}]{Mirlin2008}%
  \BibitemOpen
  \bibfield  {author} {\bibinfo {author} {\bibfnamefont {T.}~\bibnamefont
  {Ludwig}}, \bibinfo {author} {\bibfnamefont {I.~V.}\ \bibnamefont {Gornyi}},
  \bibinfo {author} {\bibfnamefont {A.~D.}\ \bibnamefont {Mirlin}}, \ and\
  \bibinfo {author} {\bibfnamefont {P.}~\bibnamefont {W\"olfle}},\ }\href
  {\doibase 10.1103/PhysRevB.77.235414} {\bibfield  {journal} {\bibinfo
  {journal} {Phys. Rev. B}\ }\textbf {\bibinfo {volume} {77}},\ \bibinfo
  {pages} {235414} (\bibinfo {year} {2008})}\BibitemShut {NoStop}%
\bibitem [{\citenamefont {Belitz}\ and\ \citenamefont
  {Kirkpatrick}(1989)}]{Belitz1989}%
  \BibitemOpen
  \bibfield  {author} {\bibinfo {author} {\bibfnamefont {D.}~\bibnamefont
  {Belitz}}\ and\ \bibinfo {author} {\bibfnamefont {T.}~\bibnamefont
  {Kirkpatrick}},\ }\href {\doibase
  https://doi.org/10.1016/0550-3213(89)90056-4} {\bibfield  {journal} {\bibinfo
   {journal} {Nuclear Physics B}\ }\textbf {\bibinfo {volume} {316}},\ \bibinfo
  {pages} {509 } (\bibinfo {year} {1989})}\BibitemShut {NoStop}%
\bibitem [{\citenamefont {Mirlin}\ and\ \citenamefont
  {W\"olfle}(1997)}]{Mirlin1997}%
  \BibitemOpen
  \bibfield  {author} {\bibinfo {author} {\bibfnamefont {A.~D.}\ \bibnamefont
  {Mirlin}}\ and\ \bibinfo {author} {\bibfnamefont {P.}~\bibnamefont
  {W\"olfle}},\ }\href {\doibase 10.1103/PhysRevB.55.5141} {\bibfield
  {journal} {\bibinfo  {journal} {Phys. Rev. B}\ }\textbf {\bibinfo {volume}
  {55}},\ \bibinfo {pages} {5141} (\bibinfo {year} {1997})}\BibitemShut
  {NoStop}%
\end{thebibliography}%
	
	\appendix

	\section{Topological term in the HLR theory\label{sec:HLR_top_term}}
	In this appendix, we derive the topological term in the gauged non-linear sigma model (NLSM) for the Halperin-Lee-Read  (HLR)\cite{Halperin1993} composite-fermion (CF) theory.
	The HLR theory differs from Son's theory\cite{Son2015} in the fact that the composite-fermion (CF) does not have a Berry phase around the Fermi surface and the emergent gauge field has a Chern-Simons term. As such, we expect the topological term to have covariant derivatives unlike the case of Son's theory which contains regular partial derivatives (see maintext section \ref{sec:top_term_derivation}). 
	
	In Refs. \onlinecite{Wang2017, Kumarsusy,KumarTrivialPaper,Huang2021}, the presence of spatially correlated potential and magnetic flux disorders was shown to be a crucial ingredient to obtain a consistent description of the half-filled Landau level in terms of the non-relativistic CFs. It was shown that the HLR CF theory exhibits emergent particle-hole (PH) symmetry and tunes the CFs at $\nu^{\rm cf}=-1$ to $\nu^{\rm cf}=0$ IQHIT transition. The correlated potential and magnetic flux disorders can be interpreted as the CFs experiencing a random magnetic field in the presence of a Zeeman term with a gyromagnetic ratio of $g=2$. In this appendix, we make use of such a disordered HLR theory in the presence of gauge fluctuations given by the following lagrangian:
	\begin{align}
		\mathcal{L} &= \eta^\dagger \left[\partial_\tau -ia_\tau+\frac{b'(\bm r)}{2m} - \mu\right.\nn\\
		&\ \ \ \ \ \ \ \ \ \ \left.  - \frac{1}{2m}(\partial_j -ia'_j(\bm r)-ia_j)^2\right]\eta + \cdots
	\end{align}
	where $\eta,\eta^\dagger$ are the annihilation and creation operators of the non-relativistic CF field, $m$ is the CF mass that controls the density of states at fermi energy $\mu$, $a_\mu$ is the fluctuating part of the emergent gauge field and $\epsilon^{ij}\partial_i a'_j(\bm r) = b'(\bm r)$ is the disordered magnetic field. The third term $b'(\bm r)\eta^\dagger \eta$ corresponds to the $g=2$ Zeeman term and generates the spatially correlated magnetic flux and chemical potential disorders. ``$\cdots$'' include additional terms such as the Chern-Simons term and gauge field kinetic energy that won't be important in this appendix.

	To derive the topological term, we add and subtract a $g=2$ Zeeman term for the fluctuating gauge field:
	\begin{align}
		\mathcal{L} &= \eta^\dagger \left[\partial_\tau -ia_\tau+\frac{b'(\bm r)+b}{2m} - \mu\right.\nn\\ 
		&\ \ \ \ \ \ \ \ \ \ \left. - \frac{1}{2m}(\partial_j -ia'_j(\bm r)-ia_j)^2\right]\eta - \frac{b \eta^\dagger\eta}{2m} + \cdots
	\end{align}
	Now, we perform the fermionic Hubbard-Stratonovic transformation of Ref. \onlinecite{Kumar2019b} and map the $g=2$ theory to a Dirac theory with a random vector potential:
	\begin{align}
		\mathcal{L} &= -\Psi^\dagger \left[  \mu +i v\sigma_j \left(\partial_j-ia'_j(\bm r)-ia_j\right)\right]\Psi \nn\\
		&\ \ \ \ \ \ \ \ \ \ - \frac{b \eta^\dagger\eta}{2m} + \eta^\dagger D^a_\tau \eta + \cdots
	\end{align}
	$\Psi \equiv (\eta\ \chi)^T$ and $\Psi^\dagger \equiv (\eta^\dagger\ \chi^\dagger)$, where $\chi,\chi^\dagger$ are the fermionic Hubbard-Stratonovic field, $D_\mu^a \equiv \partial_\mu -ia_\mu$ and $v\equiv \sqrt\frac{\mu}{2m}$. This lagrangian is nearly identical to Son's Dirac theory in the presence of vector potential disorder except for the presence of an additional Chern-Simons term for the emergent gauge field and the second to last term. Therefore, the derivation of the NLSM and the topological term is almost identical in the HLR case. A minor difference is that only the $\eta$ component of the Dirac field $\Psi$ has a time-derivative term.
	
	To see how the additional $-\frac{b \eta^\dagger\eta}{2m}$ term in the lagrangian changes the topological term, we perform disorder averaging using the replica trick and introduce the constrained NLSM field $Q$ to get:
	\begin{align}
		\mathcal{L} &= -\Psi^\dagger \left[ \mu +i v \sigma_jD_j^a + \frac{i}{2\tau}Q + \frac{b}{4m}(\sigma_z + \mathbb{1})\right]\Psi+ \cdots\label{eq:HLR_dis_avg}
	\end{align}
	where we have used the fact that $\gamma^\tau= \sigma_z$ and ignored the time-derivative term. Also, the replica indices have been suppressed for brevity. Let's integrate out fermions:
	\begin{align}
		\mathcal{L}_{\rm eff.} &= -\mathrm{Tr}\log \left[ \mu +iv\sigma_j \partial_j + \frac{v}{\sqrt{\beta}}\sigma_j\mathcal{A}_j + \frac{i}{2\tau}Q\right.\nn\\
		&\ \ \ \ \ \ \ \ \ \ \left.+ \frac{\mathcal{B}}{4m\sqrt{\beta}}(\sigma_z + \mathbb{1})\right] + \cdots
	\end{align}
	where $\mathcal{B} \equiv \epsilon^{jk}\partial_j\mathcal{A}_k$. To first order in the last term, we get:
	\begin{align}
		\mathcal{L}_{\rm eff.} &= \frac{i}{4\sqrt{\beta}}\mathrm{Tr}[Q\mathcal{B}] + \cdots\nn\\
		&= \frac{i\epsilon^{jk}}{8\sqrt{\beta}}\mathrm{Tr}[QF^{\mathcal{A}}_{ jk}]\label{eq:HLR_Hall_vertex}
	\end{align}

	If we repeat the chiral anomaly analysis of the maintext section \ref{sec:top_term_derivation} on Eq. \eqref{eq:HLR_dis_avg}, we will find that this additional term leads to the introduction of covariant derivatives in the topological term of Eq. \eqref{eq:top_term_Son}. Therefore, the HLR topological term is given by:
	\begin{gather}
		S_{\rm top} = \frac{\epsilon^{jk}}{16}\int d^2r \ \mathrm{Tr}\left[QD^a_j Q D^a_k Q\right]
	\end{gather}

	\section{Perturbation theory in gauge fluctuations at large CF conductivity\label{sec:perturbation_theory}}
	In this section, we obtain the quantum corrections to the gauge-field and the diffusive degrees of freedom that result from the coupling between them. We consider $\sigma_{xx}^{\rm cf} \gg e^2/h$ and the limit of strong Coulomb interactions. As such, the results of this appendix  strictly pertain to the experiments done at weak disorder and won't tell us much about the critical point at $\sigma_{xx}^{\rm cf} \sim \mathcal{O}(e^2/h)$. Nevertheless, we can make interesting statements about the RG flow in the regime of validity of the perturbation theory. As we'll show, the gauge fluctuations can be neglected in this limit.
	
	To perform a perturbative analysis, we expand the NLSM around the saddle point in the following way:\cite{Belitz1994}
	\begin{gather}
	Q= \; \begin{blockarray}{ccc}
	\omega > 0 & \omega < 0 & \\
	\begin{block}{(cc)c}
	\sqrt{1-V^\dagger V} & V^\dagger & \  \omega > 0 \\
	V & -\sqrt{1- VV^\dagger} & \ \omega < 0\\
	\end{block}
	\end{blockarray}\\
	Q = \sum_{n=0}^\infty Q^{(n)}\nn\\
	Q^{(0)} = \Lambda\\
	Q^{(1)} = \begin{pmatrix}
	0 & V^\dagger \\
	V & 0
	\end{pmatrix}\\
	Q^{(2)} = -\frac{{Q^{(1)}}^2\Lambda}{2}, \ \ \ Q^{(3)} = 0, \ \ \ Q^{(4)} =  -\frac{{Q^{(1)}}^4\Lambda}{8},\ \ \  \cdots
	\end{gather}
	It's useful to write the action with covariant derivatives expanded out:
	\begin{gather}
	S = S^{\rm top} +  S^{\rm a,(0)} + S^{\rm a,(1)} + S^{\rm a,(2)} \nn\\
	S^{\rm a,(0)} = \pi N_F \int d^2r\ \mathrm{Tr}\left[\frac{D}{4}(\partial_jQ)^2-\Omega Q\right] \\
	\begin{split}
	S^{\rm a,(1)} &= -\frac{i\pi \sigma_{xx}^{\rm cf}}{\sqrt{\beta}}\int d^2r\ \mathrm{Tr}\left[\mathcal{A}_j Q \partial_j Q\right]\\
	&\ \ \ \ \ \  -\frac{\pi N_F}{\sqrt{\beta}}\int d^2r\ \mathrm{Tr}[\mathcal{A}_\tau Q]
	\end{split}\label{eq:action_a_1}\\
	\begin{split}
	S^{\rm a,(2)} &= -\frac{\pi \sigma_{xx}^{\rm cf}}{2\beta}\int d^2r\ \mathrm{Tr}\left[\mathcal{A}_j Q\mathcal{A}_jQ - \mathcal{A}_j^2\right]\\
	&\ \ \ \ + \frac{N_F}{2} \sum_\alpha\int d^2rd\tau\ {(a_\tau^\alpha)}^2 \\
	&\ \ \ \ + \frac{1}{2(4\pi)^2} \sum_\alpha\int d\tau d^2rd^2r'\ U(|\bm r-\bm r'|)b^\alpha(\bm r') b^\alpha(\bm r)\label{eq:action_a_2}
	\end{split}
	\end{gather}
	The Gaussian level propagator of $Q^{(1)}$ is:
	\begin{align}
	\langle {Q^{(1)}}^{\alpha\beta}_{nm}(\bm q_1) {Q^{(1)}}^{\gamma\delta}_{kl}(-\bm q_2)\rangle_0 &= \frac{2}{\pi N_F} \frac{\Theta(-\omega_n\omega_m)}{Dq_1^2 + |\omega_n-\omega_m|}\nn\\
	& \times \delta^{(2)}(\bm q_1 - \bm q_2) \delta^{\alpha\delta}\delta^{\beta\gamma}\delta_{nl}\delta_{mk}
	\end{align}
	Notice that $\langle  Q^{(1)}(\bm q) Q^{(1)}(-\bm q) \rangle_0 \propto \frac{1}{\sigma_{xx}^{\rm cf}}$. This means that the expansion in powers of $Q^{(1)}$ is controlled by the small parameter $1/\sigma_{xx}^{\rm cf}$.
	
	Additionally, in transverse gauge, the propagator of $a_T$ for Coulomb interactions, i.e. $U(r) = e_*^2/r$ is:
	\begin{gather}
	a_j(\bm q) = \frac{i\epsilon^{jk}q_j a_T(\bm q)}{q}\\
	\langle a^\alpha_T(\omega_n,\bm q_1)a^\beta_T(-\omega_m,-\bm q_2) \rangle_0 =  \frac{8\pi}{8\pi \sigma_{xx}^{\rm cf} |\omega_n| + e_*^2 q}\nn\\
	\times \delta^{(2)}(\bm q_1-\bm q_2)\delta^{\alpha\beta}\delta_{nm}
	\end{gather}
	\subsection{Gauge-boson self-energy}
	The leading order correction to the gauge-boson propagator can be obtained to second order in the gauge coupling and $Q^{(1)}$.
	\begin{align}
	S^{\rm \Pi}[a_\mu] &\equiv \frac{1}{2} \int d^2q\  a^\alpha_\mu(\omega_n,\bm q) \Pi^{\mu\nu,\alpha\beta}_{nm}(\bm q)a^\beta_\nu(-\omega_m,-\bm q)
	\end{align}
	For the spatial part of $\Pi^{\mu\nu}$, we obtain the following term in the second order expansion in powers of the diffuson-gauge coupling term of Eq. \eqref{eq:action_a_1} and re-exponentiating the answer:
	\begin{align}
	{\Pi^{jk,\alpha\beta}_{nm}}^{(1)} &= -\sigma_{xx}^{\rm cf} \frac{|\omega_n|Dq_jq_k}{Dq^2 + |\omega_n|}\delta^{\alpha\beta}\delta_{nm}
	\end{align}
	From Eq. \eqref{eq:action_a_2}, we get the saddle point answer and two additional terms. One of them is obtained by replacing both $Q$s by $Q^{(1)}$ and the second by replacing of the $Q$s by $\Lambda$ and the other by $Q^{(2)}$. The latter term vanishes in the replica limit since $\langle Q^{(2)}\rangle_0 \propto N_r$. So, we get:
	\begin{align}
	{\Pi^{jk,\alpha\beta}_{nm}}^{(2)} &= \sigma_{xx}^{\rm cf} \delta^{\alpha\beta}\delta_{nm} \nn\\
	& - \frac{ \sigma_{xx}^{\rm cf} \delta_{n,0}\delta_{m,0}}{4\pi} \sum_{n',m'}\int d^2q'\ \frac{\Theta(-\omega_{n'}\omega_{m}')}{D{q'}^2+|\omega_{n'}-\omega_{m'}|} \label{eq:pi_2}
	\end{align}
	We can show that the second term doesn't contribute in the replica limit since we have a replica-symmetric saddle point. To see this explicitly, let's consider a simplified action of the form:
	\begin{gather}
	S[x] = \frac{1}{2}\sum_{\alpha\beta}\left(a\delta^{\alpha\beta}+b\right)x^\alpha x^\beta
	\end{gather}
	with $\alpha = 1, 2, \cdots, N_r-1, N_r$. Eigenvalues of the inverse propagator matrix are $a, a, \cdots, a, a+N_r b$. In the replica limit, all of them converge to $a$ and thus we can ignore $b$, i.e. the second term in Eq. \ref{eq:pi_2}. 
	
	Overall, we get the following for the spatial components of the polarization tensor:
	\begin{align}
	{\Pi^{jk,\alpha\beta}_{nm}}(\bm q) = \sigma_{xx}^{\rm cf}|\omega_n|\delta^{\alpha\beta}\delta_{nm} \frac{|\omega_n|\delta_{jk} + D(q^2\delta_{jk} - q_jq_k)}{Dq^2 + |\omega_n|}
	\end{align}
	Similarly:
	\begin{gather}
	{\Pi^{\tau\tau,\alpha\alpha}_{nn}}(\bm q) = N_F\frac{Dq^2}{Dq^2+|\omega_n|}\\
	{\Pi^{\tau j,\alpha\alpha}_{nn}}(\bm q) = {\Pi^{ j\tau,\alpha\alpha}_{nn}}(\bm q) = \sigma_{xx}^{\rm cf} \frac{\omega_n q_j}{Dq^2+|\omega_n|}
	\end{gather}
	where $\Pi^{\mu\nu, \alpha\beta}_{nm} = \Pi^{\mu\nu, \alpha\alpha}_{nn} \delta^{\alpha\beta}\delta_{nm}$. It can be easily verified that this satisfies the Ward-Takahashi identities $q_\mu \Pi^{\mu\nu} = \Pi^{\nu\mu}q_\mu = 0$ when we use the Fourier transform convention of Eq. \eqref{eq:Fourier_convention}. Let's convert the self-energy to transverse gauge:
	\begin{gather}
	{\Pi^{TT,\alpha\alpha}_{nn}}(\bm q) = \sigma_{xx}^{\rm cf}|\omega_n|\\
	{\Pi^{\tau T,\alpha\alpha}_{nn}}(\bm q) = {\Pi^{ T\tau,\alpha\alpha}_{nn}}(\bm q) = 0
	\end{gather}
	The first equation is the source of the $z=1$ dynamical scaling when added to the bare Coulomb term $\propto e^2_*|q| |a_T|^2$.
	
	\subsection{Diffuson self-energy: quantum corrections to conductivity}
	In principle, there are three different possible processes that can contribute to the quantum corrections to the diffusion propagator. First, since we found that $a_\tau$ is gapped at the saddle point, it mediates an effective density-density interaction with a constant form-factor which leads to the “standard” Altshuler-Aronov correction. Second, there are Hartree-type current-current gauge interactions, which in the $1/r$ case are known to be much less singular than for the short-range interaction because such processes are dominated by large transferred momenta up to $k_F$\cite{Mirlin2008}. Fortunately, resulting logarithmic corrections of both types are well under control since the absence of spin degeneracy guarantees the marginal irrelevancy of the coupling and suppression of the triplet instability\cite{Belitz1989}. Finally, one has to investigate the fate of exchange current-current diagrams involving $a_{j}$. In this subsection, we present such corrections in the NL$\sigma$M framework and demonstrate, in full agreement with Ref. \onlinecite{Mirlin1997}, that they saturate in the IR limit.

	The leading correction to the conductivity of diffusons comes from the second order expansion in powers of the action in Eq. \eqref{eq:action_a_1}. In $\beta\rightarrow \infty$ limit, we get:
	\begin{align}
	S^{Q,1}	&\sim  \frac{2 \pi\sigma_{xx}^{\rm cf} \sqrt{\pi  T \tau}}{e_*^2 \sqrt{D\tau}} \int d^2q \ Dq^2 {Q^{(1)}}^{\alpha\beta}_{nm}(\bm q) {Q^{(1)}}^{\beta\alpha}_{mn}(-\bm q) \label{eq:Diffuson_self_energy}
	\end{align}
	where we have done the momentum integral first and set the external frequencies to zero (dc limit). We then expanded in powers of $\omega_2$ and integrated over the region $\pi T <\omega_2<1/\tau$. We have ignored the constant piece that changes the zero temperature conductivity by a small amount for small $1/e_*^2$. Further, we also get the following from the same term:
	\begin{align}
	S^{Q,2} &\sim \frac{2\pi^2 \sigma_{xx}^{\rm cf}}{e_*^2  \beta\sqrt{ D}}\int d^2q \ \frac{\delta^{\alpha\beta}Dq^2}{\sqrt{|\omega_{n} - \omega_{m}|}}  \nn\\
	&\ \ \ \ \ \ \ \times   {Q^{(1)}}^{\alpha\beta}_{n,n-p}(\bm q) {Q^{(1)}}^{\beta\alpha}_{m,m+p}(-\bm q)
	\end{align}
	Notice that this is completely regular since $|\omega_{n}-\omega_{m}| \geq 2\pi /\beta$ and there is one extra sum over frequencies. 
	Similar to Ref. \onlinecite{Mirlin1997}, these terms lead to $\sqrt{T}$ dependence of the corrections to conductivity.

	In addition, we get the following corrections from the first term in Eq. \eqref{eq:action_a_2}:
	\begin{align}
	S^{Q,3} &\sim \frac{2\sigma_{xx}^{\rm cf}}{e_*^2\sqrt{D\tau}}  \int d^2q\ \left[|\omega_n-\omega_m| {Q^{(1)}}^{\alpha\beta}_{nm}(\bm q){Q^{(1)}}^{\beta\alpha}_{mn}(-\bm q)\right.\nn\\
	&\ \ \ \ \ \ \left. - \frac{\pi\delta^{\alpha\beta}}{\beta} {Q^{(1)}}^{\alpha\beta}_{n,n-p}(\bm q) {Q^{(1)}}^{\beta\alpha}_{m,m+p}(-\bm q)\right]
	\end{align}
	The first term renormalizes the coefficient of $|\omega|$ in the diffuson propagator. Since the corrections to diffusons are small, we conclude that gauge fluctuations are unimportant to this order. Also notice that these quantum corrections are controlled by the small parameter: $\sqrt{D}/e_*^2\sqrt{\tau} \propto \ell/e_*^2 \tau$, where $\ell$ is the mean-free path. 
	In terms of the renormalization group, the gauge fluctuations are irrelevant for large $\sigma_{xx}^{\rm cf}$. 
	
\end{document}